\documentclass[aps,pra,preprint,superscriptaddress,showpacs]{revtex4}

\usepackage{bm}

\usepackage{graphicx}
\usepackage{rotating}
\usepackage{epsfig}
\begin{document}
\def\G{{\cal G}}
\def\F{{\cal F}}
\def\bM{{\bf M}}
\def\bN{{\bf N}}
\def\bD{{\bf D}}
\def\bSig{{\bf \Sigma}}
\def\bLam{{\bf \Lambda}}
\def\bfeta{{\bf \eta}}
\def\a{{\bf a}}
\def\ba{{\bf a}}
\def\d{{\bf d}}
\def\P{{\bf P}}
\def\bK{{\bf K}}
\def\bk{{\bf k}}
\def\bkn{{\bf k}_{0}}
\def\bx{{\bf x}}
\def\bz{{\bf z}}
\def\bR{{\bf R}}
\def\br{{\bf r}}
\def\bq{{\bf q}}
\def\bp{{\bf p}}
\def\bG{{\bf G}}
\def\bQ{{\bf Q}}
\def\bs{{\bf s}}
\def\E{{\bf E}}
\def\bv{{\bf v}}
\def\b0{{\bf 0}}
\def\la{\langle}
\def\ra{\rangle}
\def\Im{\mathrm {Im}\;}
\def\Re{\mathrm {Re}\;}
\def\beq{\begin{equation}}
\def\eeq{\end{equation}}
\def\bea{\begin{eqnarray}}
\def\eea{\end{eqnarray}}
\def\bdm{\begin{displaymath}}
\def\edm{\end{displaymath}}
\def\bnab{{\bf \nabla}}
\def\Tr{{\mathrm{Tr}}}
\def\bJ{{\bf J}}
\def\bU{{\bf U}}
\def\bPsi{{\bf \Psi}}
\title{Pairing fluctuations and the superfluid density through the BCS-BEC
crossover}

\author{E.~Taylor}
\affiliation{Department of Physics, University of Toronto, Toronto, Ontario,
Canada, M5S 1A7,}
\author{A.~Griffin}
\affiliation{Department of Physics, University of Toronto, Toronto, Ontario,
Canada, M5S 1A7,}
\author{N.~Fukushima}
\affiliation{Institute of Physics, University of Tsukuba, Tsukuba,
Ibaraki 305, Japan,}
\author{Y.~Ohashi}
\affiliation{Institute of Physics, University of Tsukuba, Tsukuba,
Ibaraki 305, Japan,}
\affiliation{Faculty of Science and Technology, Keio University,
Hiyoshi, Yokohama, 223, Japan}
\date{\today}

\begin{abstract} 
We derive an expression for the superfluid
density of a uniform two-component Fermi gas through the BCS-BEC crossover in
terms of the thermodynamic potential in the presence of an imposed superfluid
flow.  Treating the pairing fluctuations in a Gaussian
approximation following the approach of Nozi\`eres and Schmitt-Rink, we use
this definition of $\rho_s$ to obtain an explicit result which is valid at
finite temperatures and over the full BCS-BEC crossover. It is crucial that the 
BCS gap $\Delta$, the chemical potential $\mu$, and $\rho_s$ all include the
effect of fluctuations at the same level in a self-consistent manner.  We show
that
the normal fluid density $\rho_n \equiv n - \rho_s$ naturally separates into
a sum of contributions from Fermi BCS quasiparticles ($\rho^F_{n}$) and Bose
collective modes ($\rho^B_{n}$). The expression for $\rho^F_{n}$ is
just Landau's formula for a BCS Fermi superfluid but now calculated over the
BCS-BEC crossover.  The expression for the Bose contribution $\rho^B_{n}$ is
more complicated
and only reduces to Landau's formula for a Bose superfluid in the extreme BEC
limit, where all the fermions have formed stable Bose pairs and the
Bogoliubov excitations of the associated molecular Bose condensate are
undamped. In a companion paper, we present numerical
calculations of $\rho_s$ using an expression equivalent to the one derived
in this paper, over the BCS-BEC crossover, including
unitarity, and at finite temperatures.
\end{abstract}
\maketitle
\section{Introduction}
The superfluid density $\rho_s$ is a fundamental signature in all superfluid systems~\cite{NozieresPines}.
It describes the part of the system which does not respond to an external
rotation (transverse perturbation).  
Moreover, it is an essential parameter which enters in the two-fluid
hydrodynamics of a superfluid, as first discussed by Landau in
1941~\cite{Khalatnikov}.  The superfluid density $\rho_s$ is quite different
from the
condensate density $n_c$.  In particular, it can be shown
that at $T=0$, the entire system is superfluid ($\rho_s = n$), in stark
contrast to $n_c$~\cite{Griffinbook}.  

In this paper, we define and derive an expression for the superfluid density for a
two-component Fermi gas in the BCS-BEC crossover region at finite temperatures. 
Our analysis is limited to a uniform gas.  The calculation of $\rho_s$ is based on the Leggett
mean-field BCS model of the BCS-BEC crossover, extended to include the effects
of pairing fluctuations associated with the dynamics of the bound states using
the
approach of Nozi\`eres and Schmitt-Rink (NSR)~\cite{NSR}.  The NSR
approximation has also been used to calculate the thermodynamic properties in
the BCS-BEC crossover at both $T=0$ and finite temperatures.  As shown in
detail by Hu {\it et al.}~\cite{Drummond06a,Drummond06}, this approximation
gives results that are in good agreement with Quantum Monte Carlo
calculations~\cite{Astrakharchik04,Bulgac06}.  Their work gives us confidence
in using the NSR approximation to calculate the superfluid density at finite
temperatures in the BCS-BEC crossover, apart from a small region just below
$T_c$ where the fluctuations require a more careful treatment.   

We note that in the superfluid involved with the BCS-BEC crossover, $\rho_s$ will always refer to the number of fermions which participate in the superfluid motion, not the number of Bose pairs.  Thus at $T=0$, $\rho_s=n$, where $n$ is the number density of fermions and hence, $m\rho_s$ is the total mass of the system.  

In the present paper, we define the superfluid density by imposing a ``phase twist" on the Cooper pair order parameter, 
endowing the Cooper pair condensate with a finite superfluid velocity $\bv_s$. 
Following the approach of Ref.~\cite{Fisher}, $\rho_s$ is formulated in terms 
of the second derivative of the thermodynamic potential of the superfluid with respect to
$v_s$.  We show that the normal fluid density $\rho_n \equiv n - \rho_s$ naturally separates into a sum of a Fermi
quasiparticle contribution arising from the standard BCS static mean-field
approximation plus a Bose contribution arising from
the dynamics of the pairing order parameter.  The latter contribution is treated within
a Gaussian approximation for the fluctuations around the static BCS order
parameter describing the Bose-condensed pair (Cooper)
states~\cite{Engelbrecht, Ohashi03}.  We use a single-channel model appropriate for a
broad Feshbach
resonance~\cite{Diener04}, which means that one deals with an interacting Fermi
gas with a varying $s$-wave scattering length $a_s$.  Apart from this, our
microscopic model
is identical to that used in earlier work on the collective modes in the BCS-BEC
crossover at finite temperatures~\cite{Ohashi03}.

As noted earlier, it is important to keep in mind the distinction between the
superfluid density $\rho_s$ and the condensate density $n_c$,
related to
the average occupancy of the Cooper pair state.  Numerical results for both $n_c$ and $\rho_s$ are presented in our companion
paper as a function of both $T$ and $a_s$~\cite{Fukushima}.  We note that $n_c$
has recently
been calculated at $T=0$ in the BCS-BEC crossover using a Quantum Monte Carlo simulation~\cite{Astrakharchik05}.
 
To date, $\rho_s$ in a Fermi superfluid with a Feshbach resonance has not been
calculated by such an
ab-initio method, although this has been done for superfluid
$^4$He as a function of temperature~\cite{Griffinbook,Ceperley87}.  
While it is a fundamental property of superfluids, there have been no experimental measurements of $\rho_s$ in the BCS-BEC crossover.  As we discuss briefly in the concluding section, $\rho_s$ plays a crucial role in two-fluid hydrodynamics.  This collisional domain should be accessible at finite temperatures near unitarity in the BCS-BEC crossover~\cite{Taylor05}.

We also discuss the equivalence of different formal expressions and
definitions for
the superfluid density within a given microscopic model.  We argue that relating
the normal fluid density $\rho_n$ to the thermodynamic
potential $\Omega(\bv_s)$ in the presence of a finite superfluid
flow $v_s$ gives a very elegant way of separating out the Fermi BCS
quasiparticle contribution $\rho^F_{n}$
and the Bose collective mode contribution $\rho^B_{n}$
arising from
pairing fluctuations.  When expanded out in terms of products of single-particle
BCS Green's functions (see Appendix B), our expression for $\rho^B_n$ is
extremely
complex and not physically transparent.

We show that the Fermi contribution $\rho^F_n$ to the normal fluid density
is always given by the well-known Landau formula in terms of Fermi
BCS quasiparticle excitations.  Only the values of $\Delta$ and $\mu$
appearing in the energy spectrum of these excitations change as one
sweeps through the BCS-BEC crossover.  In
the BEC limit, the fact that $\mu$ is large and negative means that the Fermi
quasiparticles are frozen out by a large effective energy gap over the
relevant temperature scale $k_BT \sim k_BT_c \ll |\mu|$ and
consequently, $\rho^F_n$ becomes negligible.

In contrast, the Bose fluctuation contribution $\rho^B_n$ to the normal fluid
density becomes increasingly dominant as we go from the BCS region to the BEC
region, where the dynamics of tightly-bound molecules dominate the
thermodynamics. Far into the BEC
region, the Bose fluctuations reduce to the usual Bogoliubov excitations
calculated in the Popov approximation, which allows for a thermal depletion of
the condensate density $n_c(T)$.  We show in detail how our general expression
for the Bose fluctuation
contribution $\rho^B_n$ to the normal fluid density reduces in the extreme BEC
limit
($|\mu| \gg k_BT_c$) to the expected Landau expression~\cite{Khalatnikov} for the normal fluid
density in terms
of undamped Bogoliubov-Popov excitations.  This reduction in the BEC limit has
recently been proven by Andrenacci, Pieri, and Strinati~\cite{Pieri} based on
a direct
diagrammatic evaluation of an expression for $\rho^B_n$ defined in terms of a
transverse velocity response function~\cite{Baym}.  However, we find that there are additional terms in our expression for $\rho_n$ which are not included in the diagrammatic analysis of Ref.~\cite{Pieri}.  These terms are
negligible in the extreme BEC limit of strongly-bound pair states, but become
important closer to unitarity where the $s$-wave scattering length $a_s$ becomes
very large. 

The present paper concentrates on the formal definition of the superfluid
density $\rho_s$ and the derivation of an explicit (but still formal) expression for a
specific microscopic model which includes
contributions from the Fermi BCS quasiparticles and the Bose pairing
fluctuations.  We concentrate on the structure of these two contributions to $\rho_n$
and the underlying 
physics of the pairing fluctuations which give rise to  $\rho^B_n$.  A companion paper by the authors~\cite{Fukushima} presents the results of extensive numerical calculations of our expression
for
$\rho_s$, as a function of
both the temperature and $s$-wave scattering length $a_s$. 
In such calculations, it is important to use 
the renormalized values of BCS gap $\Delta$ and
the Fermi chemical potential $\mu$ within a Gaussian 
approximation that includes the effects of the same pairing fluctuations which
describe the Bose collective mode contribution  $\rho^B_n$ to the normal fluid
density.

\section{Formal expression for the superfluid density}
Our expression for the superfluid density is based on the equilibrium thermodynamic potential for a current-carrying superfluid.  Thus, our starting point is the partition function 
\bea {\cal{Z}} = \int {\cal D}[\psi, \bar{\psi}] e^{-S[\psi, \bar{\psi}]}
\eea expressed as a functional integral over fermionic Grassmann
fields $\psi$ and $\bar{\psi}$~\cite{Popovbook}.  The imaginary-time action
$S[\psi,\bar{\psi}]$ is given by
\bea S[\psi, \bar{\psi}] = \int_0^\beta d\tau\left[\int
d{\bf{r}}\;\sum_{\sigma}\bar{\psi}_{\sigma}(x)\partial_{\tau}{\psi}_{\sigma}(x)
+ H\right],  \eea 
where $\beta = 1/k_BT$.  Here, we use the notation $x = ({\br, \tau})$ where
$\br$
denotes spatial
coordinates and $\tau = it$ is the imaginary time variable.  We set
$\hbar = 1$ throughout.  $H$ 
is the usual BCS pairing Hamiltonian,
\bea H &=& \int
d\br\;\sum_{\sigma}\bar{\psi}_{\sigma}(x)\left(\frac{{\hat{\bp}}^2}{2m}
-
\mu\right)\psi_{\sigma}(x) - U\int
d\br\;\bar{\psi}_{\uparrow}(x)\bar{\psi}_{\downarrow}(x){\psi}
_{\downarrow}
(x){\psi}_{\uparrow}(x).  \eea 
$U$ is the parameter characterizing the
$s$-wave
scattering interaction
between fermions in the two different hyperfine states, denoted by the
spin indices $\sigma = \uparrow,\downarrow$. From the Lippman-Schwinger
equation for the two-body scattering problem, $U$ is related to the $s$-wave
scattering length $a_s$ by~\cite{deMelo}
\bea \frac{1}{U} = -\frac{m}{4\pi a_s}
+\sum_{\bk}\left(2\varepsilon_{\bk}\right)^{-1},\eea where $\varepsilon_{\bk} =
\bk^2/2m$.  Throughout this
paper, we take the volume to be unity.  Our analysis is restricted to uniform gases.   

The Bose
pairing field $\Delta(x)$ that includes fluctuations about the mean-field
static BCS order parameter $\Delta$ is introduced through the
Hubbard-Stratonovich transformation,
\bea \lefteqn{e^{U\int d\tau\int d\br\;\bar{\psi}_{\uparrow}
\bar{\psi}_{\downarrow}{\psi}_{\downarrow}{\psi}_{\uparrow}} = }&&\nonumber\\
&&\int
{\cal{D}}[\Delta,{\Delta^*}]\exp\left\{-\int_0^\beta d\tau\int
d\br\;\left[\frac{|\Delta(x)|^2}{U}
-\left({\Delta^*(x)}{\psi}_{\downarrow}{\psi}_{\uparrow} + 
\Delta(x)\bar{\psi}_{\uparrow}\bar{\psi}_{\downarrow}\right)\right]\right\}.
\nonumber\\
\label{HS}\eea  With this identity, the partition function becomes
\bea \lefteqn{\cal{Z} =}&&\nonumber\\&&\!\!\! \int {\cal D}[\psi,
\bar{\psi}]{\cal D}[\Delta,{\Delta^*}]\exp\Bigg\{-\int_0^\beta
d\tau\int
d{\bf{r}}\;\Bigg[\sum_{\sigma}\bar{\psi}_{\sigma}(x)\left(\partial_{\tau}
+ \frac{\hat{\bp}^2}{2m} - \mu\right){\psi}_{\sigma}(x) \nonumber\\&& - 
{\Delta^*(x)}{\psi}_{\downarrow}{\psi}_{\uparrow} - 
\Delta(x)\bar{\psi}_{\uparrow}\bar{\psi}_{\downarrow} +
\frac{|\Delta(x)|^2}{U}\Bigg]\Bigg\}\nonumber\\
&=&\int {\cal D}[\psi, \bar{\psi}]{\cal
D}[\Delta,{\Delta^*}]\exp\Bigg\{-\int_0^\beta
d\tau\int
d{\bf{r}}\;\Bigg[\mathbf{\Psi}^{\dagger}\left[-\bG^{-1}\right]\mathbf{\Psi} +
\frac{|\Delta(x)|^2}{U}\Bigg]\Bigg\},\label{Z2}
\eea where we have introduced the Nambu spinors
\bea \mathbf{\Psi}^{\dagger} = \left(\bar{\psi}_{\uparrow}\;\;
\psi_{\downarrow}\right), \;\;\;\mathbf{\Psi} = \left(
\begin{array}{cc} \psi_{\uparrow}
\\ \bar{\psi}_{\downarrow}
\end{array} \right), \label{nambuspinors} \eea and $\bG^{-1}$ is the
inverse of the $2\times 2$ matrix Nambu-Gorkov BCS Green's function,
\bea {\bG^{-1}}(x,x') = \left (
\begin{array}{cc} -\partial_{\tau} - \frac{\hat{\bp}^2}{2m} + \mu &
\Delta(x)
\\
{\Delta^*}(x)
&-\partial_{\tau} + \frac{\hat{\bp}^2}{2m} - \mu 
\end{array} \right )\delta(x-x'). \label{G} 
\eea

The integration over the Grassmann fields $\psi$ in
Eq.~(\ref{Z2}) can be performed in straightforward fashion
to give
\bea {\cal {Z}} = \int {\cal D}[\Delta, {\Delta^*}]
e^{-S_{\mathrm{eff}}},\label{Bpartition}
\eea
where~\cite{deMelo}  \bea
S_{\mathrm{eff}}[\Delta, {\Delta^*}] = \int_0^\beta d\tau\int
d\br\;\frac{|\Delta(x)|^2}{U} - \mathrm{Tr}\ln [-\bG^{-1}].
\label{Seff} \eea  The trace in Eq.~(\ref{Seff}) is performed over space and
imaginary time variables, in addition to the Nambu indices.  We have used the
standard identity $\ln \mathrm{Det}\bf{A} = \mathrm{Tr}\ln\bf{A}$.  

The key function of interest in this paper is the thermodynamic potential $\Omega$, defined by
\bea\Omega = -k_BT \ln {\cal{Z}}.\label{omega0}\eea  All thermodynamic
quantities
of interest can be calculated once $\Omega$ is given in some
microscopic approximation. 
We make use of the fact that $\rho_s$ can also be obtained from
the
thermodynamic potential of a current-carrying superfluid.  To impose a current,
one applies a ``phase
twist"~\cite{Fisher} to the order parameter $\Delta(x)$:
\bea \Delta(x) \rightarrow \Delta(x) e^{i\bQ\cdot\br}. \label{phaseshift} \eea 
The
superfluid velocity $\bv_s$ associated with this imposed phase twist is
\bea \bv_s = \frac{\bQ}{M}, \eea where $M = 2m$ is the
Cooper-pair mass.  Treating $\bQ$ as small, the superfluid density is obtained from the 
lowest-order change in the free energy of the system ($F = \Omega
+ \mu N$) due
to the
added kinetic
energy of the imposed superfluid flow~\cite{Fisher}.  This extra kinetic energy is
\bea \Delta F =F(\bQ) - F(\b0) &\approx&  
\frac{Q^2}{2}\left(\frac{\partial^2 F(\bQ)}{\partial Q^2}\right)_{Q\rightarrow
0}
\equiv \frac{1}{2}\rho_s mv^2_s,
\label{changeF} \eea
with
\bea \rho_s \equiv 4m\left(\frac{\partial^2 F(\bQ)}{\partial
Q^2}\right)_{Q\rightarrow 0}. \eea  Note that the superfluid density defined
here is the superfluid number density and not the superfluid mass density used
in discussions of two-fluid hydrodynamics.  As can be seen from
Eq.~(\ref{changeF}), $\rho_sm$ is the total mass involved in the superfluid
flow, with $m$ being the Fermi atom mass.

For a fixed number of fermions $N$,
\bea \left(\frac{\partial^2 F}{\partial
Q^2}\right)_{Q\rightarrow 0} = \frac{\partial^2 \Omega}{\partial Q^2}+ 
N\frac{\partial^2\mu}{\partial Q^2}.\label{changeomega0}\eea
Microscopically, $\Omega$ can be expressed as a functional of the mean-field
gap $\Delta$, the chemical potential $\mu$, and the phase twist $Q$.  In
addition to an explicit
$Q$-dependence, $\Omega$ also depends on the phase twist implicitly through the
gap $\Delta(Q)$ and the chemical potential $\mu(Q)$.  Using these facts,
we can write Eq.~(\ref{changeomega0}) as \bea \left(\frac{\partial^2 F}{\partial
Q^2}\right)_{Q\rightarrow 0}&=&\left(\frac{\partial^2 \Omega}{\partial
Q^2}\right)_{\Delta,\mu} +
\left(\frac{\partial \Omega}{\partial \Delta}\right)_{\mu}\frac{\partial^2
\Delta}{\partial Q^2} +  
\left(\frac{\partial \Omega}{\partial \mu}\right)_{\Delta}\frac{\partial
^2\mu}{\partial Q^2}+N\frac{\partial^2\mu}{\partial Q^2}\nonumber\\
&=&\left(\frac{\partial^2 \Omega}{\partial Q^2}\right)_{\Delta,\mu}
+\left(\frac{\partial \Omega}{\partial
\Delta}\right)_{\mu}\left[\frac{\partial^2
\Delta}{\partial Q^2} -
\left(\frac{\partial\Delta}{\partial\mu}\right)
\frac{\partial^2\mu}{\partial Q^2}\right].
\label{changeomega} \eea 
In going from the first to the second line of Eq.~(\ref{changeomega}), 
we have made use of the 
number equation \bea N \equiv - \left(\frac{\partial
\Omega}{\partial\mu}\right)= - \left(\frac{\partial
\Omega}{\partial\mu}\right)_{\Delta} - \left(\frac{\partial
\Omega}{\partial\Delta}\right)_{\mu}\left(\frac{\partial
\Delta}{\partial\mu}\right).\label{NDrummond}\eea 
The evaluation of the derivatives at $Q=0$ is left implicit on the right-hand
side of Eq.~(\ref{changeomega}) and
we have also made use of the fact that the first-order corrections to $\mu$ and
$\Delta$ vanish: $(\partial \mu/\partial Q)_{Q\rightarrow 0} = (\partial \Delta
/\partial Q)_{Q\rightarrow 0} = 0$.  Separating the mean-field and fluctuation
contributions, $\Omega =
\Omega_{{\mathrm{mf}}} + \delta\Omega$, where $\Omega_{{\mathrm{mf}}} \equiv
\Omega(\Delta(x) \rightarrow \Delta)$, we can write the derivative of the
thermodynamic potential with
respect to $\Delta$ as
\bea \left(\frac{\partial \Omega}{\partial \Delta}\right)_{\mu} = 
\left(\frac{\partial \Omega_{{\mathrm{mf}}}}{\partial
\Delta}\right)_{\mu}+\left(\frac{\partial \delta\Omega}{\partial
\Delta}\right)_{\mu}. \label{dOmega}\eea The first term in Eq.~(\ref{dOmega})
vanishes since, by definition, $\partial
\Omega(\Delta(x) \rightarrow \Delta)/\partial\Delta = 0$.  As argued in
Ref.~\cite{Ohashi03}, the second term in Eq.~(\ref{dOmega}) is a higher-order
correction, beyond the Gaussian theory we use to evaluate $\Omega$ in Section
III.  Thus, for the sake of consistency we ignore this contribution and set the
second term in
Eq.~(\ref{changeomega}) equal to zero.  Our final expression for the superfluid
density is thus
\bea \rho_s = 4m\left(\frac{\partial^2\Omega(\bQ)}{\partial
Q^2}\right)_{Q\rightarrow 0} =
\frac{1}{m}\left(\frac{\partial^2\Omega(\bv_s)}{\partial
v^2_s}\right)_{v_s= 0}. \label{rhosdef}\eea
In Eq.~(\ref{rhosdef}) and elsewhere, the constancy of $\Delta$ and $\mu$
in taking derivatives with respect to $Q$ is left implicit. This formula
is the basis for our discussion of $\rho_s$ in this paper.  

We note that by ignoring terms proportional to
$(\partial\Omega/\partial\Delta)$ in Eqs.~(\ref{changeomega}) and (\ref{NDrummond}), the number equation we use to define the
superfluid density reduces to 
\bea N =  - \left(\frac{\partial
\Omega}{\partial\mu}\right)_{\Delta}. \label{N}\eea
This expression keeps
$\Delta$
fixed, meaning that derivatives of the form $(\partial\Delta/\partial\mu)$ do
not enter into the resulting equation for $N$, in contrast to
Eq.~(\ref{NDrummond}), which includes higher order corrections.  For our calculations to be consistent, the chemical potential used to evaluate our expression for
$\rho_s$ must also be calculated using Eq.~(\ref{N}), as done in 
Refs.~\cite{Engelbrecht,Ohashi03}.  The
contribution of the higher order term
$(\partial\Omega/\partial\Delta)$ to the calculation of
equilibrium thermodynamic quantities has
been discussed in some recent papers~\cite{Littlewood,Drummond06a,Drummond06}. 
In particular, in the context of the BCS-BEC crossover problem,
Refs.~\cite{Drummond06a,Drummond06} make use of the full number 
equation given by Eq.~(\ref{NDrummond}) to obtain results that are in excellent
agreement with Monte Carlo simulations at both $T=0$~\cite{Astrakharchik04} and finite $T$~\cite{Bulgac06}.  We
defer further remarks on this to Section VI.  However, it
appears from Ref.~\cite{Drummond06} that this derivative brings in the effect
of cubic and quartic fluctuations~\cite{YojiJPSJ} which have the effect of
renormalizing the strength of the effective interaction between stable Cooper
pairs~\cite{PetrovMol}.  

In Appendix A, we review the arguments demonstrating the equivalence of 
Eq.~(\ref{rhosdef}) and the usual definition of $\rho_s$ in terms of
the transverse current 
correlation function~\cite{Baym}. 

\section{thermodynamic potential for a current-carrying superfluid}

In order to calculate the thermodynamic potential for a current-carrying
superfluid, the phase twist is applied to the order parameter that enters the
inverse Green's function $\bG^{-1}$ in the action given by Eq.~(\ref{Seff}).  
To remove the phase from the order parameter, we apply the unitary transformation $\tilde{\bG}^{-1} =
\mathbf{U}^{-1}\bG^{-1}\mathbf{U}$~\cite{Ambegaokar, Stone, Thouless}, 
where 
\bea \mathbf{U} \equiv \left (
\begin{array}{cc} e^{i\bQ\cdot\br/2} & 0
\\
0
&e^{-i\bQ\cdot\br/2}
\end{array} \right ). \label{U} \eea 
Owing to the invariance of $\mathrm{Tr}\ln [-\bG^{-1}]$ with respect to the
action of a unitary
transformation of $\bG^{-1}$, 
the effective action with a
phase-twisted order parameter can be written as
\bea
S_{\mathrm{eff}}[\Delta, {\Delta^*},\bQ] = \int_0^\beta d\tau\int
d\br\;\frac{|\Delta(x)|^2}{U} - \mathrm{Tr}\ln [-\tilde{\bG}^{-1}],
\label{SeffQ} \eea where ($\hat{\bp} \equiv -i\bnab$)
\bea \tilde{\bG}^{-1}(x,x') = \left (
\begin{array}{cc} -\partial_{\tau} - \frac{(\hat{\bp}-\bQ/2)^2}{2m} + \mu &
\Delta(x)
\\
{\Delta^*}(x)
&-\partial_{\tau} + \frac{(\hat{\bp}+\bQ/2)^2}{2m} - \mu 
\end{array} \right )\delta(x-x'). \label{Gtilde} \eea 

The thermodynamic potential for a current-carring superfluid can
be evaluated from this action, using
Eqs.~(\ref{Bpartition}) and (\ref{omega0}), once  some approximation is introduced so that the functional integration in 
Eq.~(\ref{Bpartition}) can be carried out. 
Following the standard prescription,
we expand the action in powers of fluctuation about the mean-field BCS pairing
field:
$\Delta(x) =
\Delta
+ \Lambda(x)$; $\tilde{\bG}^{-1}
= \tilde{\bG}_0^{-1} + \bSig$, where $\tilde{\bG}_0^{-1} =
\tilde{\bG}^{-1}(\Delta(x) \rightarrow \Delta)$ 
and \bea \bSig = \left (
\begin{array}{cc} 0 & \Lambda(x)
\\
\bar{\Lambda}(x)
&0
\end{array} \right )\delta(x-x'). \label{Sigma}
\eea  
Clearly $\Lambda(x)$ corresponds to the fermionic self-energies due to coupling
to Bose
collective modes involving pair fluctuations in the Cooper pair channel. 

Using the expansion
$\mathrm{Tr}\ln[-\tilde{\bG}^{-1}] = \mathrm{Tr}\ln[-\tilde{\bG}^{-1}_0(1
+\tilde{\bG}_0\tilde{\bSig})] = \mathrm{Tr}\ln[-\tilde{\bG}^{-1}_0]  + 
\mathrm{Tr}\ln[1+\tilde{\bG}_0\tilde{\bSig}] =
\mathrm{Tr}\ln[-\tilde{\bG}^{-1}_0] + 
\sum_{n=1}\mathrm{Tr}[(\tilde{\bG}_0\tilde{\bSig})^n](-1)^{n+1}/n$, we expand
Eq.~(\ref{SeffQ}) up to quadratic order in
the Bose fluctuation field $\Lambda$ to obtain
the Gaussian action, $S_{\mathrm{Gauss}} \equiv S^{(0)} + S^{(2)}$. 
Fourier-transforming, the 
mean-field $S^{(0)}$ and fluctuation $S^{(2)}$ contributions are given by
\bea S^{(0)} = \beta  \frac{\Delta^{2}}{U} - \sum_{k}\mathrm{tr}\ln
[-\tilde{\bG}_0^{-1}(k)] \label{S0k}\eea
and 
\bea S^{(2)} &=& \beta \sum_k\frac{|\Lambda_k|^2}{U} +
\frac{1}{2}\sum_{k,q}\mathrm{tr}[\tilde{\bG}_0(k)\bSig(-q)
\tilde{\bG}_0(k+q)\bSig(q)]
\nonumber\\ &\equiv&
\frac{1}{2}\sum_q\bLam^{\dagger} \tilde{\bM} \bLam.
\label{S2kb} \eea In Eq.~(\ref{S2kb}), $q\equiv (\bq,i\nu_m)$ and $k\equiv
(\bk,i\omega_m)$ are 4-vectors denoting the momenta
$\bq$ and $\bk$ as well as the Bose and Fermi Matsubara frequencies $i\nu_m =
2\pi m/\beta$ and $i\omega_n =
2\pi(n+1)/\beta$, respectively, where $m,n$ are integers.  In momentum-frequency
space, the mean-field (denoted by the subscript ``0") Nambu-Gorkov BCS Green's
function $\tilde{\bG}_0(k)$ for the current-carrying BCS superfluid is defined
by its inverse, 
\bea \tilde{\bG}^{-1}_0(k) = \left (i\omega_n -\frac{\bk\cdot\bQ}{2m}\right)-
\left(\xi_{\bk} + \frac{Q^2}{8m}\right)\hat{\tau}_3 + \Delta\hat{\tau}_1.
\label{G0Q} \eea
Here, $\xi_{\bk} \equiv \bk^2/2m - \mu$, while $\hat{\tau}_1,\hat{\tau}_3$ are
Pauli
spin matrices.
We have assumed that the mean-field order
parameter $\Delta = \Delta^*$ is real.  In the last line of
Eq.~(\ref{S2kb}), we have defined the spinor 
$\bLam^{\dagger} \equiv (\bar{\Lambda}(q), \Lambda(-q))$,
and the matrix elements of the inverse $2\times2$ matrix pair fluctuation
propagator $\tilde{\bM}$
for a current-carrying superfluid are given by
\bea \frac{\tilde{M}_{11}(q)}{\beta} =
\frac{\tilde{M}_{22}(-q)}{\beta }=\frac{1}{U} +
\frac{1}{\beta}\sum_k \tilde{G}_{0,11}(k+q)\tilde{G}_{0,22}(k)
\label{m11Q}\eea
and
\bea \frac{\tilde{M}_{12}(q)}{\beta} =
\frac{\tilde{M}_{21}(q)}{\beta}=\frac{1}{\beta}\sum_k
\tilde{G}_{0,12}(k+q)\tilde{G}_{0,12}(k).\label{m12Q}\eea 
Here, $\tilde{G}_{0,ij}$ denotes the $ij$-th element of the matrix mean-field BCS
Green's function defined by Eq.~(\ref{G0Q}). 

Substituting $S_{\mathrm{eff}} \approx  S^{(0)} + S^{(2)}$
into
Eq.~(\ref{Bpartition}) and performing the Gaussian 
integration over the Bose fluctuation fields $(\bar{\Lambda},\Lambda)$, the
thermodynamic potential for a current-carrying superfluid reduces to \bea \Omega(\bQ) &=& \frac{\Delta^{2}}{U} -
\frac{1}{\beta}\sum_{k}\mathrm{tr}\ln
[-\tilde{\bG}^{-1}_0(k)]+\frac{1}{2\beta}\sum_{q}\ln {\mathrm{det}}
\tilde{\bM}(q)\nonumber\\
&\equiv&\Omega^F(\bQ) + \Omega^B(\bQ).\label{OmegaQ}\eea
This formula will be used to calculate
$\rho_s$ in Eq.~(\ref{rhosdef}) and thus plays a key role in the rest of this
paper.  The first two terms of Eq.~(\ref{OmegaQ})
comprise the mean-field contribution from Fermi BCS quasiparticles,
\bea \Omega^F(\bQ)&=& \frac{\Delta^{2}}{U} -
\frac{1}{\beta}\sum_{k}\mathrm{tr}\ln
[-\tilde{\bG}^{-1}_0(k)].\label{OmegaF}\eea
We
emphasize that the
values of
$\Delta$ and $\mu$ in Eq.~(\ref{G0Q}) evaluated using our Gaussian theory are strongly
renormalized from their mean-field values by the effects of fluctuations in the
Cooper 
pair field and the associated collective modes, as described by the NSR
theory~\cite{NSR,Engelbrecht,Ohashi03}.    The values of these microscopic parameters for a current-carrying superfluid are obtained by self-consistently solving the gap equation, $(\partial  S^{(0)}/\partial \Delta) = 0$, together with the number equation $N = -(\Omega(\bQ)/\partial\mu)_{\Delta}$, where $S^{(0)}$ is given by Eq.~(\ref{S0k}) and $\Omega(\bQ)$ is given by Eq.~(\ref{OmegaQ}).  
Recall that our expression for the superfluid density, given by Eq.~(\ref{rhosdef}), leaves $\Delta$ and $\mu$ fixed, so we only require the values of these quantities in the current-free state, found from  $\partial  S^{(0)}(\bQ=\b0)/\partial \Delta = 0$ and $N = -(\Omega(\b0)/\partial\mu)_{\Delta}$.  Further details of this calculation are given in Ref.~\cite{Fukushima}.

The contribution from the Bose
collective
modes in Eq.~(\ref{OmegaQ}) is 
\bea \Omega^B(\bQ)&=&\frac{1}{2\beta}\sum_{\bq,i\nu_m}\ln {\mathrm{det}}
\tilde{\bM}(\bq,i\nu_m). \label{OmegaB}\eea
The collective modes will be shown to play an
increasingly important
role in $\rho_n$ as one goes from the BCS to the BEC regimes.
The spectrum $\omega_{\bq}$ of the collective modes is determined from
\bea \mathrm{det}\mathbf{\tilde{M}}(\bq,i\nu_m \rightarrow
\omega_{\bq} + i0^+) =
\tilde{M}_{11}(q)\tilde{M}_{11}(-q)-\tilde{M}^2_{12}(q)=0,\label{spectrum}\eea
where $i\nu_m \rightarrow
\omega_{\bq}+ i0^+$ denotes the usual analytic continuation from imaginary
Bose frequencies.  In most of the BCS-BEC crossover, these collective modes will be damped at finite temperatures (i.e., $\omega_{\bq}$ has an imaginary part).

\section{superfluid density in the BCS-BEC crossover}
In this section, we derive an explicit expression for the superfluid density
in the crossover starting from the formula given by Eq.~(\ref{rhosdef}) for
the model defined in Section III. From
Eqs.~(\ref{G0Q}) and (\ref{OmegaQ}), one sees
that the thermodynamic potential for a superfluid 
with a finite superfluid velocity $\bv_s = \bQ/M$ is equivalent to the 
thermodynamic potential for a current-free superfluid ($\bv_s = \b0$),
but where
the chemical potential and Matsubara frequencies are now
Doppler-shifted~\cite{Stone}: \bea \mu \rightarrow \mu - Q^2/8m\equiv
\tilde{\mu},\label{mushift}\eea \bea
i\omega_n \rightarrow i\omega_n -
\bk\cdot\bQ/2m\equiv i\tilde{\omega}_n, \eea \bea i\nu_m \rightarrow i\nu_m
-
\bq\cdot\bQ/2m\equiv i\tilde{\nu}_m.\eea
Considering separately the effects of
the shifts to the chemical potential and the Matsubara
frequencies, we can write the second-order
derivative of $\Omega$ with respect
to $Q$ (keeping $\Delta$ and $\mu$ fixed) as 
\bea \frac{\partial^2\Omega}{\partial Q^2} &=& \left(\frac{\partial^2
\tilde{\mu}}{\partial Q^2}\right)\frac{\partial\Omega}{\partial \tilde{\mu}} +
2\left(\frac{\partial \tilde{\mu}}{\partial
Q}\right)\frac{\partial^2\Omega}{\partial
\tilde{\mu}\partial Q}
+\left(\frac{\partial^2\Omega}{\partial Q^2}\right)_{\tilde{\mu}}\nonumber\\
&=&-\frac{1}{4m}\frac{\partial\Omega}{\partial \tilde{\mu}} 
-\frac{Q}{2m}\frac{\partial^2\Omega}{\partial
\tilde{\mu}\partial Q}
+\left(\frac{\partial^2\Omega}{\partial Q^2}\right)_{\tilde{\mu}}.\label{d2dQ2}
\eea
Evaluated at
$Q=0$, the middle
term
in Eq.~(\ref{d2dQ2}) vanishes and Eq.~(\ref{rhosdef}) reduces to
\bea \rho_s &=& -\left(\frac{\partial \Omega}{\partial
\tilde{\mu}}\right)_{Q\rightarrow 0} +
4m\left(\frac{\partial^2\Omega}{\partial Q^2}\right)_{\tilde{\mu},
Q\rightarrow 0}\nonumber\\
&=&n + 4m\left(\frac{\partial^2\Omega}{\partial
Q^2}\right)_{\tilde{\mu}, Q\rightarrow 0}.  \label{rhosdef2}\eea
In the last line, we have made use of the number equation $n =
-(\partial\Omega/\partial\tilde{\mu})_{\mu,\Delta,Q\rightarrow 0} =
-(\partial\Omega/\partial\mu)_{\Delta}$.  Since $n \equiv \rho_s + \rho_n$,
Eq.~(\ref{rhosdef2})
gives us the following expression for the normal fluid density:
\bea \rho_n = -4m\left(\frac{\partial^2\Omega}{\partial
Q^2}\right)_{\tilde{\mu}, Q\rightarrow 0}.  \label{rhondef}\eea

Carrying out the summation over Fermi Matsubara frequencies in
Eq.~(\ref{OmegaF}), the mean-field BCS quasiparticle contribution to
the
thermodynamic potential in the presence of a current becomes~\cite{QFT}
\bea \Omega^F(\bQ) &=& \frac{\Delta^{2}}{U} +
\sum_{\bk}\left(\tilde{\xi}_{\bk}-\tilde{E}_{\bk}\right)-
\frac{2}{\beta}\sum_{\bk}\ln\left[1 +
e^{-\beta(\bk\cdot\bQ/2m +\tilde{E}_{\bk})}\right],\label{OmegaFsimp}\eea
where the single-particle quasiparticle energies are given by $\tilde{E}_{\bk} =
\sqrt{\tilde{\xi}^2_{\bk} + \Delta^2}$ with $\xi_{\bk} \equiv \bk^2/2m -
\tilde{\mu}$, where $\tilde{\mu}$ is the Doppler-shifted chemical potential defined in
Eq.~(\ref{mushift}).  

Summing over the fermion Matsubara frequencies in Eqs.~(\ref{m11Q}) and
(\ref{m12Q}), the matrix elements of the inverse matrix propagator
for pair fluctuations in the
current-carrying superfluid are given by
\bea \tilde{M}_{11}(q) &=& \tilde{M}_{22}(-q) = \frac{\beta}{U}
+\sum_{\bk}\Bigg[\left(f^+_{\bk}
- f^-_{\bk + \bq}\right)
\frac{v^2_{\bk}v^2_{\bk
+ \bq}}{i\nu_m - \bq\cdot\bQ/2m + \tilde{E}_{\bk} + \tilde{E}_{\bk +
\bq}}\nonumber\\
&&+\left(f^-_{\bk} - f^+_{\bk + \bq}\right) \frac{u^2_{\bk}u^2_{\bk +
\bq}}{i\nu_m - \bq\cdot\bQ/2m - \tilde{E}_{\bk} - \tilde{E}_{\bk +
\bq}}\nonumber\\
&&+\left(f^+_{\bk} - f^+_{\bk + \bq}\right) \frac{v^2_{\bk}u^2_{\bk +
\bq}}{i\nu_m - \bq\cdot\bQ/2m + \tilde{E}_{\bk} - \tilde{E}_{\bk +
\bq}}\nonumber\\
&&+\left(f^-_{\bk} - f^-_{\bk + \bq}\right) \frac{u^2_{\bk}v^2_{\bk +
\bq}}{i\nu_m - \bq\cdot\bQ/2m - \tilde{E}_{\bk} + \tilde{E}_{\bk + \bq}}
\Bigg] \label{m11b}\eea
and
\bea \tilde{M}_{12}(q) &=&  \tilde{M}_{21}(q)=\sum_{\bk}\Bigg[\left(f^-_{\bk+
\bq} -
f^+_{\bk}\right) \frac{u_{\bk}v_{\bk}u_{\bk + \bq}v_{\bk + \bq}}{i\nu_m -
\bq\cdot\bQ/2m + \tilde{E}_{\bk} + \tilde{E}_{\bk + \bq}}\nonumber\\
&&+\left(f^+_{\bk+\bq} - f^-_{\bk}\right) \frac{u_{\bk}v_{\bk}u_{\bk +
\bq}v_{\bk + \bq}}{i\nu_m - \bq\cdot\bQ/2m - \tilde{E}_{\bk} - \tilde{E}_{\bk +
\bq}}\nonumber\\
&&+\left(f^+_{\bk} - f^+_{\bk + \bq}\right) \frac{u_{\bk}v_{\bk}u_{\bk +
\bq}v_{\bk + \bq}}{i\nu_m - \bq\cdot\bQ/2m + \tilde{E}_{\bk} - \tilde{E}_{\bk +
\bq}}\nonumber\\
&&+\left(f^-_{\bk} - f^-_{\bk + \bq}\right) \frac{u_{\bk}v_{\bk}u_{\bk +
\bq}v_{\bk + \bq}}{i\nu_m - \bq\cdot\bQ/2m - \tilde{E}_{\bk} + \tilde{E}_{\bk +
\bq}}\Bigg], \label{m12b}\eea
where
\bea f^{\pm}_{\bp} \equiv f\left(\bp\cdot\bQ/2m \pm \tilde{E}_{\bp}\right) \eea
are
the Fermi
distribution functions. 
Here, $u_{\bp} = \sqrt{(1 + \tilde{\xi}_{\bp}/\tilde{E}_{\bp})/2}$ and  $v_{\bp}
= \sqrt{(1 -
\tilde{\xi}_{\bp}/\tilde{E}_{\bp})/2}$ are the usual Bogoliubov quasiparticle
amplitudes.  Recall that the normal fluid density
is evaluated at fixed $\tilde{\mu}$ and consequently, the 
dependence of $\tilde{E}_{\bp}$ on $Q$ can
be ignored for the sake of calculating $\rho_n$ in Eq.~(\ref{rhondef}).  The
expressions given by Eqs.~(\ref{m11b}) and (\ref{m12b}) reduce to the
standard
expressions in the literature~\cite{Engelbrecht} for $M_{ij}(\bq,i\nu_m)$ when 
$\bv_s = \b0$. 

The distribution functions appearing in Eqs.~(\ref{m11b}) and
(\ref{m12b}) involve Doppler-shifted Fermi quasiparticle energies:
$\bp\cdot\bQ/2m
\pm \tilde{E}_{\bp}$.  The shift $\bp\cdot\bQ/2m$ reflects the fact that
additional
Fermi
quasiparticles will be excited when the superfluid velocity is finite since
thermal equilibrium is defined with respect to the stationary lab
frame~\cite{Stone}.

Using the thermodynamic potential in
Eq.~(\ref{OmegaQ}), the normal fluid density $\rho_n$ is given by the sum of
Fermi quasiparticle and Bose collective mode contributions:
\bea \rho_n = \rho^F_n + \rho^B_n,\label{rhoBFn}\eea where
\bea \rho^F_n
&=&-\frac{m}{\beta}\sum_k \left(\frac{\bk\cdot\hat{\bQ}}{m}\right)^2
\mathrm{tr}[\bG_0(k)\bG_0(k)] \label{rhoFn} \eea
and
\bea \rho^B_n =-\frac{2m}{\beta}\sum_q
\frac{1}{\left(\mathrm{det}\tilde{\bM}\right)^2}\left[\mathrm{det}
\tilde{\bM}\left(\frac{\partial^2\;\mathrm{det}
\tilde{\bM}}{\partial Q^2}\right)_{\tilde{\mu}}-\left(\frac{\partial\;
\mathrm{det}\tilde{\bM}}{\partial
Q}\right)^2_{\tilde{\mu}} \right]_{Q\rightarrow 0}. \label{rhoBn}\eea  Here,
$\hat{\bQ} =
\bQ/|\bQ|$.  The expression for the Bose contribution $\rho^B_n$ is very
compactly given in terms of the
determinant of the inverse fluctuation propagator $\mathrm{det}\tilde{\bM}$,
the zeros of which give the spectrum of the Bose collective modes.
The simplicity of this expression for $\rho^B_n$ is lost 
when expanded in terms of products of
current-free BCS Green's
function (see Appendix B).  One can show after a
little work that the result given by Eq.~(\ref{rhoBn}) is identical to that
obtained in Ref.~\cite{Fukushima} based on a calculation of the current response
to a superfluid flow.  

The normal fluid density $\rho^F_n$ due to
Fermi
BCS quasiparticles given in Eq.~(\ref{rhoFn})
is readily identified as the long-wavelength, static
limit ($q
\rightarrow 0$) of the BCS current-current correlation function (multiplied by
$-m$).   
Carrying out the Matsubara frequency sum in the usual way, Eq.~(\ref{rhoFn})
reduces to 
%To carry out the summation over
%fermion Matsubara frequencies in Eq.~(\ref{trGG}), we use the standard
%formula (see, for instance, Ref.~\cite{Mahan}),
%$1/\beta\sum_{i\omega_n}f(i\omega_n) =
%(-)\sum_{z_i}\mathrm{Res}[f(z)n_{f(b)}(z),z_i]$, for fermions (bosons) where
%$z_i$ denotes the poles of some function
%$f(z)$, $\mathrm{Res}[f(z)n_{f(b)}(z), z_i]$  is the residue of
%$f(z)n_{f(b)}(z)$
%evaluated at $z_i$, and $n_{f(b)}$ is the Fermi (Bose)
%distribution function.  Since the poles appearing in Eq.~(\ref{trGG})
%are
%second-order, we
%also use the identity \bea \mathrm{Res}[F(z), z_i] = \lim_{z\rightarrow z_i}
%\frac{d}{dz}[(z-z_i)^2 F(z)]\label{residue2}\eea to calculate the residue of
%$F(z) = f(z)f(z)$ at the second-order pole $z_i$.  Using
%Eq.~(\ref{residue2}) to evaluate the frequency sum and 
%passing into the continuum limit, Eq.~(\ref{trGG}) becomes
\bea \rho^F_{n}&=&-\frac{2}{m}\int
\frac{d^3\bk}{(2\pi)^3} (\bk\cdot\hat{\bQ})^2 \frac{\partial
f(E_{\bk})}{\partial E_{\bk}}
\nonumber\\
&=& \frac{2}{3m}\int \frac{d^3\bk}{(2\pi)^3} \bk^2\left(-\frac{\partial
f(E_{\bk})}{\partial E_{\bk}}\right).
\label{landauF}  \eea
This is the well-known Landau formula for
the normal fluid density of a uniform weak-coupling BCS superfluid, arising from
thermally-excited 
Fermi BCS quasiparticles~\cite{statphys2}.  In our case, it is
valid for the entire BCS-BEC crossover, taking into account that the
quasiparticle spectrum depends on 
$\Delta$ and $\mu$ which are renormalized from their mean-field BCS values
by the inclusion of the effects of Bose fluctuations~\cite{Drummond06,Engelbrecht,Ohashi03}.  Note that the Landau formula given by 
Eq.~(\ref{landauF}) also results by using Eq.~(\ref{OmegaFsimp}) in Eq.~(\ref{rhondef}).

Eq.~(\ref{rhoBn}) describes the contributions to the normal
fluid
density from
fluctuations  $\delta\Delta$ of the Bose pairing field.  
In general, the Bose pair excitations are damped at finite temperatures,
coupling to the 
continuum of BCS quasiparticle
states. As a result, the Bose fluctuations will have a finite lifetime and
$\rho^B_{n}$ will not reduce to the usual Landau formula involving
Bose excitations.  In the BEC limit, however, the
pair binding energy becomes very large and BCS quasiparticles are strongly
suppressed.  As a result, damping will not occur.  In
this limit, we expect that our expression for $\rho^B_{n}$ will be given by
Landau's formula for a Bose superfluid.  In the next section, we give the details of this proof.

\section{The normal fluid density in the BEC limit}

Close to unitarity and on the BCS side of the BCS-BEC crossover, Landau
damping of the Bose collective modes described in Section IV arises due to scattering processes that
involve BCS
Fermi quasiparticles: $\tilde{\omega}_{\bq} + \tilde{E}_{\bk} =
\tilde{E}_{\bk+\bq}$. 
When such damping occurs, the collective
modes
strongly hybridize with BCS Fermi quasiparticles and the concept of
well-defined, long-lived Bose excitations breaks down.  In this region, we do
not expect the normal fluid density  $\rho^B_{n}$ to be given by a Landau formula for the Bose excitations.  In the strong-coupling BEC limit, however, the fermions form                   
bound pairs with a large binding energy~\cite{deMelo}:
$E_{\mathrm{binding}} = -1/ma^2_s$.  As a result of this large binding energy,
Fermi quasiparticle excitations, which involve the breakup of pairs, become completely
frozen out over the experimentally relevant temperature scale $k_BT \sim k_BT_c
\ll E_{\mathrm{binding}}$.  Because the Fermi quasiparticles are
frozen out, they no longer contribute to Landau damping of the
Bose collective modes.  Thus, 
in the BEC limit, the normal fluid is comprised of a gas of well-defined Bose
excitations and one expects $\rho^B_n$ will reduce to the usual Landau expression
for Bose excitations in this limit~\cite{Fetter}.   In this section, we show how this result emerges from our formalism (which is valid in the entire BCS-BEC crossover) in the BEC limit.

Deep in the BEC region, the chemical potential becomes increasingly large and
negative.  In the strong-coupling limit where $\Delta,
Q^2/8m,k_BT \ll |\mu|$, the BCS gap equation for the
current-carrying superfluid ($Q = Mv_s$),
\bea \frac{\Delta}{U} =
\frac{1}{\beta}\sum_k\tilde{G}_{0,12}(k),\label{gap}\eea
can be solved analytically.  This gives $\tilde{\mu} \equiv \mu - 
Q^2/8m = -1/(2ma^2_s)$, which is one-half the molecular binding
energy~\cite{deMelo}. 
When
$|\tilde{\mu}| \gg k_B T$, the BCS
quasiparticles are frozen out ($f^{+}_{\bp} \rightarrow 0$;
$f^{-}_{\bp} \rightarrow 
1$) and the BCS quasiparticle contribution $\rho^F_n$, given by Eq.~(\ref{landauF}),
vanishes.

In the low-energy regime $\omega_{\bq} \ll |\tilde{\mu}|$, the spectrum of Bose 
excitations is expected to have the form $\sqrt{(c \,\bq)^2 + (\bq^2/2m^{*})^2}$. 
To extract the contribution of these modes to the normal fluid density
$\rho^B_{n}$,
we set $f^+ = 0$ and $f^-= 1$ in Eqs.~(\ref{m11Q}) and (\ref{m12Q}) and then
expand the
inverse fluctuation propagator matrix elements in powers of $q$.  This procedure gives
\bea \frac{\tilde{M}_{11}(q)}{\beta} \simeq A + B|\bq|^2 + C(i\nu_m -
\bq\cdot\bQ/2m)^2 + D(i\nu_m -
\bq\cdot\bQ/2m), \label{m11e}\eea
and 
\bea \frac{\tilde{M}_{12}(q)}{\beta} \simeq A + F|\bq|^2 + G(i\nu_m -
\bq\cdot\bQ/2m)^2.\label{m12e}\eea  We note that outside the BEC region, where
$k_BT \sim
{\cal{O}}(|\tilde{\mu}|)$, we cannot set $f^+ = 0$, $f^-= 1$. 
Consequently,
the terms in
the inverse fluctuation propagator responsible for Landau damping, given by
the last two lines in Eq.~(\ref{m11b}) and Eq.~(\ref{m12b}), cannot be
neglected.  
In this case, it is well-known that one cannot carry out an expansion
in powers of $\bq$ and
$i\tilde{\nu}_m$, as in Eqs.~(\ref{m11e}) and (\ref{m12e}), since these terms
are
singular in the long wavelength, zero frequency limit~\cite{tsuneto,stoof}. 
This means that the expansions in Eqs.~(\ref{m11e}) and (\ref{m12e}) are not valid
in the unitarity or BCS regions.

Apart from the shift to the chemical potential given by Eq.~(\ref{mushift}), the
expansion coefficients in Eqs.~(\ref{m11e}) and
(\ref{m12e}) are the same as for the $Q=0$ case given
in Ref.~\cite{Engelbrecht}, namely
\bea A = \sum_{\bk}\frac{\Delta^2}{4\tilde{E}^3_{\bk}}, \eea
\bea B = \sum_{\bk}\left[\left(2 -
3\frac{\Delta^2}{\tilde{E}^2_{\bk}}\right)\frac{\tilde{\xi}_{\bk}}{m} +
\frac{|\bk|^2\cos^2\phi}{m^2}\left(-2 + 13\frac{\Delta^2}{\tilde{E}^2_{\bk}} -
10\frac{\Delta^4}{\tilde{E}^4_{\bk}}\right)\right]\frac{1}{16\tilde{E}^3_{\bk}
},
\eea
\bea C = \sum_{\bk}\left(\frac{\Delta^2}{\tilde{E}^2_{\bk}} -
2\right)\frac{1}{16\tilde{E}^3_{\bk}}, \eea
\bea D = -\sum_{\bk}\frac{\tilde{\xi}_{\bk}}{4\tilde{E}^3_{\bk}}, \eea
\bea F = \sum_{\bk}\left[-3\frac{\Delta^2}{\tilde{E}^2_{\bk}}
\frac{\tilde{\xi}_{\bk}}{m} +
\frac{|\bk|^2\cos^2\phi}{m^2}\left(7\frac{\Delta^2}{\tilde{E}^2_{\bk}} -
10\frac{\Delta^4_0}{\tilde{E}^4_{\bk}}\right)\right]\frac{1}{16\tilde{E}^3_{\bk}
},
\eea
and
\bea G =
\sum_{\bk}\left(\frac{\Delta^2}{\tilde{E}^2_{\bk}}\right)
\frac{1}{16\tilde{E}^3_{\bk}}. \eea We have made use of the gap equation, 
given by Eq.~(\ref{gap}), to eliminate $1/U$ from
$(\beta)^{-1}\tilde{M}_{11}(q)$.  In the
strong-coupling BEC limit, $\Delta\ll |\tilde{\mu}|$, and we can further expand the
integrands in powers of $\Delta/|\tilde{\mu}|$.  To leading order, using
$|\tilde{\mu}| =
(2ma^2_s)^{-1}$, we find
\bea A \approx \Delta^2\sum_{\bk}\frac{1}{4\tilde{\xi}^3_{\bk}} =
\frac{\Delta^2 a^3_s m^3}{16\pi},\label{A}\eea
\bea B \approx \sum_{\bk}\left[\frac{1}{8m\tilde{\xi}^2_{\bk}} -
\frac{|\bk|^2\cos^2\phi}{4m^2\tilde{\xi}^3_{\bk}}\right] =
\frac{ma_s}{32\pi},\label{B}\eea
\bea C \approx -\sum_{\bk}\frac{1}{8\tilde{\xi}^3_{\bk}} =
-\frac{m^3a^3_s}{16\pi}, \label{C}\eea and
\bea D \approx -\sum_{\bk}\frac{1}{4\tilde{\xi}^2_{\bk}} = -\frac{m^2
a_s}{8\pi}.\label{D}\eea
To leading order, we find $F\sim \Delta^2 a^5_s$ and $G\sim
\Delta^2a^7_s$, which are 
vanishingly small in the BEC limit, $a_s\rightarrow 0$. Similarly, since $C \propto (ma_s)^3$,
we set this coefficient equal to zero as
well.  However, since $\Delta^2 \propto a^{-1}_s$, one finds that $A \propto a^2_s$, and
we retain
$A$ in Eqs.~(\ref{m11e}) and (\ref{m12e}).

With 
coefficients given by Eqs.~(\ref{A}), (\ref{B}), and (\ref{D}), and
setting $C=F=G=0$, we find
\bea \mathrm{det}\tilde{\bM}(\bq, i\tilde{\nu}_m) =
2AB\bq^2 + B^2\bq^4 - D^2\left(i\nu_m - \frac{\bq\cdot\bQ}{M}\right)^2.
\label{spectrum2}\eea
Since the fluctuations spectrum is given by the zeros of
$\mathrm{det}\tilde{\bM}(\bq, \omega_{\bq})$, one finds
\bea \omega_{\bq}(\bv_s) &=& \bq\cdot\bv_s + 
\sqrt{\frac{2AB}{D^2}\bq^2 +
\frac{B^2}{D^2}\bq^4} \nonumber\\
&=&\bq\cdot\bv_s + \sqrt{\frac{\Delta^2a^2_s}{4}\bq^2 +
\left(\frac{\bq^2}{2M}\right)^2}.\label{dispersion}\eea
We note that the value of $\Delta$ appearing in this expression is 
temperature-dependent.  Since $\Delta(T)\neq 0$ is
associated with the existence of a molecular Bose condensate in the BEC region of interest,  the dispersion of Bose collective modes can be written in terms of the condensate
density $n_c$.  In Ref.~\cite{Fukushima}, we show that the corrections $\delta n_c$
to the mean-field expression for the condensate density,  \bea n_{c0}(T)
= \sum_{\bk}\frac{\Delta^{2}(T)}{4E^2_{\bk}}\tanh^2(\beta
E_{\bk}/2),\label{nc}\eea are negligible throughout the BCS-BEC crossover, within
our NSR Gaussian approximation. 
Thus, we can use Eq.~(\ref{nc}) to determine the condensate density in
the BEC limit (where $|\mu|\gg k_BT$), \bea
n_{c}(T)
= \frac{\Delta^2(T)M^2 a_s}{32\pi}.\label{ncBEC}\eea  It is important to
emphasize that in obtaining this expression, we have only taken the
limit $|\mu|/k_BT \rightarrow \infty$, where $\tanh^2(\beta
E_{\bk}/2)\rightarrow 1$.  However, $\Delta$ still has a strong temperature
dependence arising from the thermally excited pairing fluctuations which are
not
frozen out.  The temperature dependence of $\Delta(T)$ is calculated
within our Gaussian approximation throughout the BCS-BEC crossover in
Ref.~\cite{Fukushima}. 

Using the result in Eq.~(\ref{ncBEC}), one can show that the sound velocity in
Eq.~(\ref{dispersion}) can be written as
\bea c^2 \equiv \frac{\Delta(T)a^2_s}{4} = \frac{U_Mn_{c}(T)}{M}\label{c}\eea
for an interacting gas of bosons of mass $M=2m$. This is the standard
Bogoliubov-Popov sound velocity with $U_M = 4\pi a_M/M$, but with the molecular
scattering length given by the mean-field result $a_{M} =
2a_s$~\cite{deMelo}.  

In order to get the correct value of the molecular scattering length $a_M\simeq
0.6a_s$
in the BEC limit~\cite{PetrovMol}, one would have to include the effects of
4-body correlations
which are beyond the 2-body physics contained in our Gaussian theory;
i.e., we would need to 
expand the action to quartic order in fluctuations~\cite{YojiJPSJ}.  As pointed
out by Hu {\it et al.}~\cite{Drummond06}, the correct renormalized value of $a_M$
emerges when one calculates $\mu$ using the number equation given in
Eq.~(\ref{NDrummond}) that includes the contribution from
$\partial\Omega/\partial\Delta$.  Thus, while we do not
consider it in this paper, it appears that we understand how our
present calculation can be improved to get the correct value of $a_M\simeq 0.6a_s$.   

Using the expression for $\mathrm{det}\tilde{\bM}(\bq,i\tilde{\nu}_m)$
given by 
Eq.~(\ref{spectrum2}), it is straightforward to evaluate $\rho^B_n$
in Eq.~(\ref{rhoBn}).
Making use of Eq.~(\ref{dispersion}), Eq.~(\ref{spectrum2}) reduces to
\bea  \mathrm{det}\tilde{\bM}(\bq,i\tilde{\nu}_m) = -D^2\left[\left(i\nu_m - \bq\cdot\bQ/M\right)^2 - 
\omega^2_{\bq}(\bv_s=\b0)\right]. \label{detMBEC}\eea Using this expression, we
find\bea
\lefteqn{\frac{1}{\left(\mathrm{det}\tilde{\bM}\right)^2}\left[\mathrm{det}
\tilde{\bM}\left(\frac{\partial^2\;\mathrm{det}
\tilde{\bM}}{\partial Q^2}\right)_{\tilde{\mu}}-\left(\frac{\partial\;
\mathrm{det}\tilde{\bM}}{\partial
Q}\right)^2_{\tilde{\mu}} \right]_{Q\rightarrow 0}=}\nonumber\\
&&\frac{1}{D^4\left[(i\nu_m)^2 - \omega^2_{\bq}\right]^2}
\left\{2D^4\left[(i\nu_m)^2 -
\omega^2_{\bq}\right]\left(\frac{\bq\cdot\hat{\bQ}}{M}\right)^2
- 4D^4(i\nu_m)^2\left(\frac{\bq\cdot\hat{\bQ}}{M}\right)^2\right\},
\eea where $\omega_{\bq} = \omega_{\bq}(Q=0)$ is the usual
Bogoliubov-Popov excitation energy in the absence of a superfluid flow, $\bv_s = \b0$.
Using this result in Eq.~(\ref{rhoBn}), and recalling that $\rho^F_n$ vanishes
in
the BEC limit, we obtain
\bea \rho_n = \rho^B_{n}&=& \frac{M}{\beta}\sum_{\bq,i\nu_m}
\left(
\frac{\bq\cdot\hat{\bQ}}{M}\right)^2\frac{2(i\nu_m)^2 +
2\omega_{\bq}^2}{(i\nu_m - \omega_{\bq})^2(i\nu_m +
\omega_{\bq})^2}.\label{rhos3.5}\eea
To bring out the physics of Eq.~(\ref{rhos3.5}), it can also be written in
terms of
the transverse 
current correlation function
for a dilute Bose gas of interacting molecules~\cite{Fetter},\bea
\rho^B_{n}&=&\frac{M}{\beta}\sum_{\bq,i\nu_m}
\left(\frac{\bq\cdot\hat{\bQ}}{M}\right)^2
\mathrm{tr}\left[\bD(\bq,i\nu_m)\bD(\bq,i\nu_m)\right], \label{rhos4}\eea
where 
\bea \bD(\bq,i\nu_m) = \frac{1}{(i\nu_m)^2 -\omega_{\bq}^2}\left (
\begin{array}{cc} i\nu_m + \omega_{\bq}&0
\\
0
&i\nu_m -\omega_{\bq}
\end{array} \right ) \label{D0} \eea is the $2\times 2$ Bose propagator 
describing the
Bogoliubov excitations.
Carrying out the Bose frequency sum in Eq.~(\ref{rhos4}) as in
Ref.~\cite{Fetter}, we find the
expected result
\bea \rho^B_{n}
&=& -\frac{2}{M}\int \frac{d^3\bq}{(2\pi)^3}(\bq\cdot\hat{\bQ})^2\frac{\partial
n_B(\omega_{\bq})}{\partial \omega_{\bq}}
\nonumber\\
&=& \frac{2}{3M}\int \frac{d^3\bq}{(2\pi)^3}
\bq^2\left(-\frac{\partial n_B(\omega_{\bq})}{\partial \omega_{\bq}}\right),
\label{rhos5}  \eea where $n_B(\omega) = (e^{\beta \omega} - 1)^{-1}$ is the    
Bose distribution function. 
Equation~(\ref{rhos5}) is precisely Landau's formula for the
normal fluid density of a Bose gas described in terms of Bogoliubov
excitations~\cite{Fetter}.   Recall from Section I that $\rho_s$ and hence
$\rho_n$ always refers to the number of fermions.  Thus, Eq.~(\ref{rhos5}) is
twice the usual expression~\cite{Fetter}, reflecting the fact that it is
counting the number of fermions (not the number of bosons) involved with a
normal fluid composed of Bogoliubov excitations of a molecular BEC.  

As we have already discussed, retaining terms in Eq.~(\ref{NDrummond}) that are
proportional to $(\partial\Omega/\partial\Delta)$ leads to the
renormalization of the molecular scattering length, from $a_M=2a_s$ to $a_M \simeq 
0.6a_s$.  To be consistent, one must include the analogous terms in
Eq.~(\ref{changeomega}) and  
additional terms will be generated in
our definition of the superfluid density given by Eq.~(\ref{rhosdef}):
$\rho_s\rightarrow\rho_s +
4m(\partial\Omega/\partial\Delta)_{\mu}[\partial^2\Delta/\partial Q^2 - 
(\partial\Delta/\partial\mu)\partial^2\mu/\partial Q^2]$.
In the extreme BEC
limit, however, $\partial^2\Delta/\partial Q^2\rightarrow 0$ as the BCS
quasiparticles become
frozen out, and
$\partial^2\mu/\partial
Q^2\rightarrow 1/4m$ [as shown below Eq.~(\ref{gap})] so that $\rho_s
\rightarrow \rho_s + n_{pf,\Delta}$, where $n_{pf,\Delta} \equiv
(\partial\Omega/\partial\Delta)_{\mu}(\partial\Delta/\partial\mu)$ is the
correction to the number equation~\cite{Drummond06}.  Using this new expression
in
Eq.~(\ref{rhosdef2}), $n_{pf,\Delta}$ just adds another contribution to
the total density $n$ and Eq.~(\ref{rhondef}) remains unchanged in the
BEC limit. Thus,
even if we retain terms in Eqs.~(\ref{changeomega}) and (\ref{NDrummond}) that
lead to the
renormalization of the molecular scattering length, in the BEC limit, $\rho_n$
is still given by Eq.~(\ref{rhondef}).  Consequently, our major result in Eq.~(\ref{rhos5}) still holds in the BEC limit when we include the higher-order corrections,
except that
$a_M$ will now be $\simeq0.6a_s$. 

For completeness, we write down the pair fluctuation contribution to the
thermodynamic potential in Eq.~(\ref{OmegaB}) in the BEC limit.  Using
Eq.~(\ref{detMBEC}), the Bose Matsubara frequency sum can be evaluated
analytically~\cite{QFT} and we find
\bea \Omega^B(\bQ) &=&
\frac{1}{2}\sum_{\bq}\omega_{\bq}+
\frac{1}{\beta}\sum_{\bq}\ln\left[1-e^{-\beta(
\bq\cdot\bQ/M + \omega_{\bq})}\right].\label{OmegaBBec}\eea
Inserting Eq.~(\ref{OmegaBBec}) into Eq.~(\ref{rhondef}) also leads to
Eq.~(\ref{rhos5}).   The integrands in Eq.~(\ref{OmegaBBec}) are
strictly only valid at small momenta, such that $\bq^2 /2M \ll \Delta$, and the
divergent zero-point energy contribution to Eq.~(\ref{OmegaBBec}) must be
regularized by an appropriate choice of cutoff. 

The result given in Eq.~(\ref{rhos5}) for the BEC limit of the BCS-BEC
expression has also been derived in Ref.~\cite{Pieri} 
using a diagrammatic approach.  This is discussed in Appendix B.

\section{Conclusions}

In the present paper, we have derived an explicit formula for the normal fluid
density $\rho_n$ in terms
of two
contributions.  One is the expected contribution $\rho^F_n$ given by
Eq.~(\ref{rhoFn}) arising from Fermi BCS single-particle excitations.  In the
BEC limit,
$\rho^F_n$ vanishes since the effective quasiparticle energy gap becomes very
large.  Physically, the pair binding energy becomes very large and the
Fermi quasiparticles, which are excitations corresponding to the breakup of these
pair states, become frozen out.  

The most interesting contribution to $\rho_n$ in the BCS-BEC crossover is the
contribution $\rho^B_n$ from collective modes associated with the dynamics of
the pair states.  This is given in our NSR formalism by 
Eq.~(\ref{rhoBn}).  Within this Gaussian approximation to the pair fluctuation
propagator, as summarized in Eqs.~(\ref{m11b}) and (\ref{m12b}), one can
proceed to calculate $\rho^B_n$ numerically.  The results of such calculations
are discussed in a companion paper~\cite{Fukushima} over
the whole BCS-BEC crossover and as a function of
temperature.  

The delicate nature of the Cooper pair molecule in the unitarity region of the
crossover leads to damping of the collective modes given by
Eq.~(\ref{spectrum}).  This means that, in general, $\rho^B_n$ is not given by
a simple Landau expression such as Eq.~(\ref{rhos5}).   However, one does
expect such a Landau formula to emerge in the extreme BEC limit where the
Cooper pairs become very strongly bound and the system reduces to a weakly
interacting Bose gas of stable molecules.  In Section V, we showed how this
expected result emerges naturally from our general formalism.

In deriving our key starting formula for the superfluid density given by
Eq.~(\ref{rhosdef}), we neglected the contribution
of $(\partial\Omega/\partial\Delta)$ in Eqs.~(\ref{changeomega}) and
(\ref{NDrummond}). We argued that such a term is a
higher order correction which cannot be
consistently included in a Gaussian approximation on which our formal 
analysis and numerical calculation~\cite{Fukushima} are based.  The role
of the second term
in Eq.~(\ref{NDrummond}) has been discussed in recent
calculations~\cite{Littlewood,Drummond06a,Drummond06}.  In particular, Hu, Lui,
and Drummond~\cite{Drummond06} have shown that in the BEC limit, this term in
the number equation gives rise to a renormalization of the molecular scattering
length from the mean-field value $a_M = 2a_s$ to the correct value  $a_M\simeq 0.6a_s$.  This result is consistent with the calculation of
Ohashi~\cite{YojiJPSJ} who went past our NSR Gaussian pairing fluctuation
approximation to include the effects of cubic and
quartic fluctuations (for a diagrammatic analysis, see
Refs.~\cite{Strinati00,Strinati05,Brodsky06}). Ohashi found that these higher
order effects lead to a
renormalization of the effective interaction between molecules, and obtained a
value for $a_M$ close to the result of Petrov {\it et al.}~\cite{PetrovMol}.

We conclude that the neglected contribution to
Eq.~(\ref{NDrummond}) picks up an important class of fluctuations
left out of our Gaussian model, which are precisely those needed to give
the correct molecular scattering length in the BEC limit.  As one knows from other problems, derivatives of the Gaussian thermodynamic potential can generate results which describe an improved model.  This emphasizes the usefulness of calculating $\rho_s$ starting from the result in Eq.~(\ref{rhosdef}).

The superfluid density was first introduced by Landau in connection with a
two-fluid theory for the collisional hydrodynamics of a Bose superfluid~\cite{Khalatnikov}.  The
form of the Landau two-fluid hydrodynamics is generic for any superfluid with a
two-component order parameter (amplitude and phase)~\cite{Taylor05}.  The
frequencies of the
resulting
hydrodynamic modes are given completely in terms of the equilibrium
thermodynamic functions, including the superfluid density.  The precise values
of these equilibrium quantities depend on the nature of the dominant thermal
excitations, which can be different in different superfluids. In the BCS-BEC
crossover, one goes from the BCS limit, where Fermi BCS quasiparticles dominate
the thermodynamics, to the BEC limit where the Bose collective modes
(Bogoliubov-Popov excitations) dominate the thermodynamics.  

As a result, a
careful discussion of the two-fluid collective modes requires a careful analysis
of the changing weights of the Fermi and Bose excitations as we pass
through the BCS-BEC crossover, both for thermodynamic quantities such as the
entropy and compressibility as well as the equilibrium superfluid density.  The
advantage of
calculating $\rho_s$ from the second derivative of the thermodynamic potential
$\Omega(\bv_s)$ calculated within a Gaussian approximation for the
fluctuations, as we do in our work (see also Ref.~\cite{Fukushima}), is that
all other thermodynamic functions can also be determined from
$\Omega(\bv_s=\b0)$. Heiselberg~\cite{Heiselberg06} has given an informative
first
study of first and second
sound in the BCS-BEC crossover for a uniform gas by calculating $\rho_s$ and
other thermodynamic functions in the BEC and BCS limits and interpolating into
the unitarity region ($|a_s|\rightarrow \infty$).  We hope to give a more
definitive
discussion of first and second sound using the numerical results for
$\rho_s$ given in Ref.~\cite{Fukushima}.

\begin{acknowledgments}
E.T. would like to thank Pierbiagio Pieri and Menderes Iskin for helpful
discussions. A.G. and E.T. were supported by NSERC of Canada.   Y.O. was
financially supported by Grant-in-Aid
for Scientific research from the Ministry of Education, Culture, Sports,
Science and Technology of Japan (16740187, 17540368, and 18043005).  
\end{acknowledgments}

\appendix

\section{Superfluid density and the transverse current correlation function}
In this section we review the relationship between our definition of the
superfluid density and the transverse response definition commonly evoked in the
literature~\cite{Baym}.  We start with the partition function expressed in
terms of both
Bose and
Fermi fields, given by Eq.~(\ref{Z2}).  Applying a phase twist to the
order parameter as was done in Section III, Eq.~(\ref{Z2}) is written as
\bea {\cal{Z}} = \int {\cal{D}}[\bar{\psi},\psi]{\cal{D}}[{\Delta^*},\Delta]
e^{-S[\bar{\psi},\psi,{\Delta^*},\Delta,\bQ]}, \eea
where
\bea S[\bar{\psi},\psi,{\Delta^*},\Delta,\bQ] =
\int\;d^4x\;\left[\bPsi^{\dagger}(x)\left[-\tilde{\bG}^{-1}(x,x')\right]\bPsi
(x')+
\frac{|\Delta|^2}{U}\right],\label{Sdelferm}\eea
and $\bG^{-1}(x,x')$ is given by Eq.~(\ref{Gtilde}).  From Eq.~(\ref{Gtilde}),
keeping $\Delta$ and $\mu$ fixed, 
\bea\frac{\partial \tilde{\bG}^{-1}(x,x')}{\partial Q}=
\left(\frac{\hat{\bp}\cdot
\hat{\bQ}}{2m} -
\frac{Q}{4m}
\hat{\tau}_3\right)\delta(x-x')\label{dGdQ}\eea
and
\bea \frac{\partial^2 \tilde{\bG}^{-1}(x,x')}{\partial Q^2}= -\left(
\frac{1}{4m}
\hat{\tau}_3\right)\delta(x-x').\label{d2GdQ2}\eea
Using Eqs.~(\ref{Sdelferm}), (\ref{dGdQ}), and (\ref{d2GdQ2}), we obtain the
relations
\bea \frac{\partial S}{\partial Q} =
\int\;d^4x\;\bPsi^{\dagger}(x)\left[\left(-\frac{\hat{\bp}\cdot\hat{\bQ}}{2m} +
\frac{Q}{4m}
\hat{\tau}_3\right)\delta(x-x')\right]\bPsi(x') \label{dSdQ}\eea
and
\bea \frac{\partial^2 S}{\partial Q^2} =
\int\;d^4x\;\bPsi^{\dagger}(x)\left[\left(
\frac{1}{4m}
\hat{\tau}_3\right)\delta(x-x')\right]\bPsi(x'). \label{d2SdQ2}\eea
We use these expressions to obtain the relation
\bea \frac{\partial^2}{\partial Q^2}e^{-S}\Bigg|_{Q\rightarrow 0}&=&\left[\left(
\frac{\partial S}{\partial Q}\right)^2 - \frac{\partial^2 S}{\partial Q^2} 
\right]e^{-S}\Bigg|_{Q
=0}\nonumber\\
&=&-\Bigg[\frac{1}{4m}\int d^4x\;\bPsi^{\dagger}\hat{\tau}_3\bPsi  \nonumber\\
&&-\int d^4x\;d^4x'\;\left(\bPsi^{\dagger}(x)\frac{\hat{\bp}\cdot\hat{\bQ}}{2m}
\bPsi(x)\right)\left(\bPsi^{\dagger}(x')\frac{\hat{\bp}'\cdot\hat{\bQ}}{2m}
\bPsi(x')\right)\Bigg]e^{-S}\nonumber\\
&=&-\left(\frac{\beta}{4m}\hat{N}
-\frac{1}{4}\int d^4x\;d^4x'\;\hat{j}_z(x)\hat{j}_z(x')\right)e^{-S},
\label{d2SdQ2app}\eea
where
\bea \hat{N} = \int
d\br\;\sum_{\sigma}\bar{\Psi}_{\sigma}(\br)\Psi_{\sigma}(\br)\eea
is the number operator, and, having arbitrarily chosen $\hat{\bQ} = \hat{\bz}$,
\bea
\hat{j}_z = \frac{1}{2mi}\sum_{\sigma}
\left(\bar{\Psi}_{\sigma}\left(\frac{\partial}{\partial z}\Psi_{\sigma}\right)
- 
\left(\frac{\partial}{\partial z}\bar{\Psi}_{\sigma}\right)\Psi_{\sigma}\right)
\eea is the $z$-component of the current density operator.

Using Eq.~(\ref{d2SdQ2app}) along with the thermodynamic potential in the
presence of a superfluid flow, \bea
\Omega(\bQ) = -T\ln\int {\cal{D}}[\bar{\psi},\psi]{\cal{D}}[{\Delta^*},\Delta]
e^{-S[\bar{\psi},\psi,{\Delta^*},\Delta,\bQ]},
\eea
we obtain \bea
\frac{\partial^2\Omega}{\partial Q^2}\Bigg|_{Q\rightarrow 0}&=&
-\frac{T}{\cal{Z}}
\int
{\cal{D}}[\bar{\psi},\psi]{\cal{D}}[{\Delta^*},\Delta]
\frac{\partial^2}{\partial
Q^2}e^{-S}\Bigg|_{Q\rightarrow 0} \nonumber\\&&-
T\left(\frac{1}{\cal{Z}}\int{\cal{D}}
[\bar{\psi},\psi]{\cal{D}}[{\Delta^*},\Delta]
\left(\frac{\partial S}{\partial Q}\right)e^{-S}\Bigg|_{Q\rightarrow 0}\right)^2
\nonumber\\
&=&\frac{1}{\cal Z}_0\int
{\cal{D}}[\bar{\psi},\psi]{\cal{D}}[{\Delta^*},\Delta]
\left(\frac{1}{4m}\hat{N}
-\frac{1}{4\beta}\int d^4x\;d^4x'\;\hat{j}_z(x)\hat{j}_z(x')\right)e^{-S}
\nonumber\\
&=& \frac{N}{4m} - \frac{1}{4}\langle
\hat{J}_z\hat{J}_z\rangle_0,\label{JJ}\eea
where $N = \langle \hat{N}\rangle_0$ and $\hat{J}_z \equiv \beta^{-1/2}\int
d^4x\;\hat{j}_z$.   
$\langle \cdots\rangle_0$ denotes an expectation value with respect to 
the current-free state: \bea
\langle \cdots \rangle \equiv \frac{1}{\cal{Z}}_0\int{\cal{D}}
[\bar{\psi},\psi]{\cal{D}}[{\Delta^*},\Delta](\cdots)
e^{-S[\bar{\psi},\psi,{\Delta^*},\Delta,\bQ=\b0]},
\eea
and
\bea 
{\cal{Z}}_0 = \int{\cal{D}}
[\bar{\psi},\psi]{\cal{D}}[{\Delta^*},\Delta]
e^{-S[\bar{\psi},\psi,{\Delta^*},\Delta,\bQ=\b0]}.\eea  
Note that the second line in Eq.~(\ref{JJ}) vanishes by symmetry, i.e., $\langle
\hat{J}_z
\rangle_0 = 0$.  

Comparing Eq.~(\ref{JJ}) with Eq.~(\ref{rhosdef}), we obtain the result
\bea \rho_s = n - m\langle
\hat{J}_z\hat{J}_z\rangle_0, \eea which identifies~\cite{Baym}
\bea \rho_n = m\langle
\hat{J}_z\hat{J}_z\rangle_0\eea as the normal fluid density.

\section{normal fluid contribution from the Bose fluctuations}

A central result of our paper is the formal expression in Eq.~(\ref{rhoBn}) for
the Bose fluctuation contribution to the normal fluid density, where the
$\tilde{\bM}$ matrix elements are given in
Eqs.~(\ref{m11b}) and (\ref{m12b}).  In this appendix, we ``unpack" this formal
result for $\rho^B_n$ to give it more explicitly in terms of single-particle
Nambu-Gorkov Green's functions of a {\it current-free} BCS superfluid.  This
will allow us to compare with other results in the literature~\cite{Pieri}.  

Expanding the $Q$-derivatives in Eq.~(\ref{rhoBn}), the Bose fluctuation
contribution to $\rho_n$ is given by
\bea \rho^B_n
&=&\frac{2m}{\beta}\sum_q
\frac{1}{\left(M_{11}M_{22}-M_{12}M_{12}\right)^2}\Bigg[M_{11}M_{11}
\frac{\partial \tilde{M}_{22}}{\partial Q}
\frac{\partial\tilde{M}_{22}}{\partial
Q}\nonumber\\
&&-4M_{11}M_{12}\frac{\partial \tilde{M}_{22}}{\partial Q}
\frac{\partial \tilde{M}_{12}}{\partial Q} + 
2M_{11}M_{22}\frac{\partial \tilde{M}_{12}}{\partial Q}
\frac{\partial \tilde{M}_{12}}{\partial Q}+2M_{12}M_{12}\frac{\partial
\tilde{M}_{11}}{\partial Q}
\frac{\partial \tilde{M}_{22}}{\partial Q}
\nonumber\\&&+2M_{12}M_{12}\frac{\partial \tilde{M}_{12}}{\partial Q}
\frac{\partial \tilde{M}_{12}}{\partial Q} - 4
M_{22}M_{12}\frac{\partial \tilde{M}_{11}}{\partial Q}
\frac{\partial \tilde{M}_{12}}{\partial Q}
+M_{22}M_{22}\frac{\partial \tilde{M}_{11}}{\partial Q}
\frac{\partial \tilde{M}_{11}}{\partial Q}\nonumber\\
&&
-(M_{11}M_{22}-M_{12}M_{12})\left(M_{22}\frac{\partial^2\tilde{M}_{11}}{\partial
Q^2}+M_{11}\frac{\partial^2\tilde{M}_{22}}{\partial
Q^2} - 2M_{12}\frac{\partial^2\tilde{M}_{12}}{\partial
Q^2}\right)\Bigg]_{\tilde{\mu},Q\rightarrow 0}.
\label{rhosexp}\eea Here, $M_{ij} = \tilde{M}_{ij}(\bQ=\b0)$.  
To express Eq.~(\ref{rhosexp}) in terms of current-free Green's functions,
we use the identities
\bea \frac{\partial \tilde{G}_{0,11}(p)}{\partial
Q}\Bigg|_{\tilde{\mu},Q\rightarrow 0} =
\left(\frac{\bp\cdot\hat{\bQ}}{2m}\right)\left[G_{0,11}(p)G_{0,11}(p) +
G_{0,12}(p)G_{0,12}(p)\right],\eea
\bea \frac{\partial \tilde{G}_{0,12}(p)}{\partial
Q}\Bigg|_{\tilde{\mu},Q\rightarrow 0} =
\left(\frac{\bp\cdot\hat{\bQ}}{2m}\right)\left[G_{0,12}(p)G_{0,11}(p) +
G_{0,12}(p)G_{0,22}(p)\right],\eea
\bea \frac{\partial \tilde{G}_{0,22}(p)}{\partial
Q}\Bigg|_{\tilde{\mu},Q\rightarrow 0} =
\left(\frac{\bp\cdot\hat{\bQ}}{2m}\right)\left[G_{0,22}(p)G_{0,22}(p) +
G_{0,12}(p)G_{0,12}(p)\right],\eea
\bea \lefteqn{\frac{\partial^2 \tilde{G}_{0,11}(p)}{\partial
Q^2}\Bigg|_{\tilde{\mu},Q\rightarrow 0} =}&&\nonumber\\
&&2\left(\frac{\bp\cdot\hat{\bQ}}{2m}\right)^2\left[G_{0,11}(p)\left(
G^2_{0,11}(p) + G^2_{0,12}(p)\right) + 
G^2_{0,12}(p)\left(G_{0,11}(p) +G_{0,22}(p)\right)\right],\eea
\bea \lefteqn{\frac{\partial^2 \tilde{G}_{0,12}(p)}{\partial
Q^2}\Bigg|_{\tilde{\mu},Q\rightarrow 0} =}&&\nonumber\\
&&2\left(\frac{\bp\cdot\hat{\bQ}}{2m}\right)^2G_{0,12}(p)\left[
G^2_{0,11}(p) + G^2_{0,22}(p)
+G^2_{0,12}(p) +G_{0,11}(p)G_{0,22}(p)\right],\eea
and
\bea \lefteqn{\frac{\partial^2 \tilde{G}_{0,22}(p)}{\partial
Q^2}\Bigg|_{\tilde{\mu},Q\rightarrow 0} =}&&\nonumber\\
&&2\left(\frac{\bp\cdot\hat{\bQ}}{2m}\right)^2\left[G_{0,22}(p)\left
(
G^2_{0,22}(p) + G^2_{0,12}(p)\right) + 
G^2_{0,12}(p)\left(G_{0,11}(p) +G_{0,22}(p)\right)\right].\eea
With these identities and substituting the BCS gap equation given by
Eq.~(\ref{gap})
into Eq.~(\ref{m11Q}), we find
\bea \lefteqn{\frac{\partial \tilde{M}_{11}(q)}{\partial
Q}\Bigg|_{\tilde{\mu},Q\rightarrow 0} =\frac{\partial
\tilde{M}^{22}(-q)}{\partial
Q}\Bigg|_{\tilde{\mu},Q\rightarrow 0}=\sum_{k}\left(\frac{\bk\cdot\hat{\bQ}
+ 
\bq\cdot\hat{\bQ}}{m}\right)}&&\nonumber\\
&&\times\Big[G_{0,11}(k+q)G_{0,11}(k+q)
+G_{0,12}(k+q)G_{0,12}(k+q)\Big]G_{0,22}(k),\label{m11Qder1}\eea
\bea  \lefteqn{\frac{\partial \tilde{M}_{12}(q)}{\partial
Q}\Bigg|_{\tilde{\mu},Q\rightarrow 0} =\sum_{k}\left(\frac{\bk\cdot\hat{\bQ}
+ 
\bq\cdot\hat{\bQ}}{m}\right)}&&\nonumber\\
&&\times\Big[G_{0,12}(k+q)G_{0,11}(k+q)
+G_{0,12}(k+q)G_{0,22}(k+q)\Big]G_{0,12}(k)
,\label{m12Qder1}\eea
\bea \lefteqn{\frac{\partial^2 \tilde{M}_{11}(q)}{\partial
Q^2}\Bigg|_{\tilde{\mu},Q\rightarrow 0} =\frac{\partial^2
\tilde{M}_{22}(-q)}{\partial
Q^2}\Bigg|_{\tilde{\mu},Q\rightarrow 0}=}&&\nonumber\\
&&2\sum_k\left(\frac{\bk\cdot\hat{\bQ}}{2m}\right)^2
\frac{1}{\Delta}\Bigg\{G_{0,11}(k)G_{0,12}(k)\left[G_{0,11}(k)+
G_{0,22}(k)\right]\nonumber\\
&&\;\;\;\;\;+G_{0,12}(k)\left[G_{0,12}(k)G_{0,12}(k) +
G_{0,22}(k)G_{0,22}(k)\right]\Bigg\}\nonumber\\
&&+2\sum_{k}\Bigg\{2\left(\frac{(\bk\cdot\hat{\bQ}
+ 
\bq\cdot\hat{\bQ})}{2m}\right)^2\Bigg[G_{0,22}(k)G_{0,11}(k+q)\Big[
G_{0,11}(k+q)G_{0,11}(k+q) \nonumber\\&&\;\;\;+
G_{0,12}(k+q)G_{0,12}(k+q)\Big]+G_{0,22}(k)G_{0,12}(k+q)G_{0,12}(k+q)\Big[
G_{0,11}(k+q) +
G_{0,22}(k+q)\Big]\Bigg]\nonumber\\&&+\frac{(\bk\cdot\hat{\bQ}
+ 
\bq\cdot\hat{\bQ})(\bk\cdot\hat{\bQ})}{2m^2}
\Big[
G_{0,11}(k+q)G_{0,11}(k+q) +
G_{0,12}(k+q)G_{0,12}(k+q)\Big]\nonumber\\&&
\;\;\;\times\Big[
G_{0,22}(k)G_{0,22}(k) + G_{0,12}(k)G_{0,12}(k)\Big] \Bigg\},
\label{m11Qder2}\eea
and
\bea \lefteqn{\frac{\partial^2 \tilde{M}_{12}(q)}{\partial
Q^2}\Bigg|_{\tilde{\mu},Q\rightarrow 0} =}&&\nonumber\\&&
+2\sum_{k}\Bigg\{2\left(\frac{(\bk\cdot\hat{\bQ}
+ 
\bq\cdot\hat{\bQ})}{2m}\right)^2G_{0,12}(k)G_{0,12}(k+q)\Big[
G_{0,11}(k+q)G_{0,11}(k+q) \nonumber\\&&\;\;\;+
G_{0,22}(k+q)G_{0,22}(k+q) + G_{0,12}(k+q)G_{0,12}(k+q) +
G_{0,11}(k+q)G_{0,22}(k+q)\Big]\nonumber\\&&+\frac{(\bk\cdot\hat{\bQ}
+ 
\bq\cdot\hat{\bQ})(\bk\cdot\hat{\bQ})}{2m^2}
G_{0,12}(k+q)\Big[
G_{0,11}(k+q) + G_{0,22}(k+q)\Big]\nonumber\\&&
\;\;\;\times G_{0,12}(k)\Big[
G_{0,11}(k) + G_{0,22}(k)\Big]\Bigg\}.
\label{m12Qder2}\eea
In deriving Eqs.~(\ref{m11Qder1})-(\ref{m12Qder2}), we have made use of the following
identities for BCS Green's functions:
\bea G_{0,11}(-k) = -G_{0,22}(k) \label{G11relation}\eea
and
\bea G_{0,12}(-k) = G_{0,12}(k).\label{G12relation}\eea
Taken together, Eqs.~(\ref{rhosexp}) and (\ref{m11Qder1})-(\ref{m12Qder2})
give an explicit expression for the normal fluid density due to Bose
fluctuations in terms of products
of BCS Green's functions.

It is of interest to relate our result for $\rho^B_n$ to that obtained recently by
Andrenacci {\it et
al.}~\cite{Pieri}, who used a direct diagrammatic evaluation of the response
function definition of the normal fluid density (see Appendix
A). 
Eqs.~(14) and (25) in Ref.~\cite{Pieri} give the following expression for the
fluctuation contribution to the normal fluid density:
\bea \rho^B_{n,AL} = -m\chi^{AL}_{z,z}(Q=0),\label{rhonAL} \eea
where
\bea \chi^{AL}_{z,z}(Q=0) &=&
-\frac{1}{(2m)^2}\frac{1}{\beta^3}\sum_{k,k',q}\sum_{i,i',i''}\sum_{j,j',j''}(2k
_z + 2q_z)(2k'_z +
2q_z)\Gamma_{j',i'}(q)\Gamma(q)_{i'',j''}(q)\nonumber\\&& \times
G_{0,ii'}(k+q)
G_{0,i''i}(k+q)G_{0,i''i'}(-k)G_{0,j'j}(k'+q)
G_{0,jj''}(k'+q)\nonumber\\
&&\times G_{0,j'j''}(-k'). \label{AL}\eea 
This result is based on the Aslamazov-Larkin-type (AL) diagrammatic
contributions to the
transverse current
correlation function $\chi_{z,z}$.  Using our notation as defined in the text of
this paper, the vertex functions in Eq.~(\ref{AL}) are 
\bea \Gamma_{11}(q) = \Gamma_{22}(-q)=\beta
\frac{M_{22}(q)}{\mathrm{det}\bM},\label{Gamma11}\eea
and
\bea \Gamma_{12}(q) = \Gamma_{21}(q) = \beta
\frac{M_{12}(q)}{\mathrm{det}\bM}.\label{Gamma12}\eea
To facilitate comparison with our results for $\rho^B_n$, we
expand Eqs.~(\ref{rhonAL}) and (\ref{AL}) to
give 
\bea \rho^B_{n,AL} &=&\frac{1}{m\beta^3}\sum_{k,k',q}(k_z + q_z)(k'_z +
q_z)\times\nonumber\\&&\Bigg\{\Gamma_{11}(q)\Gamma_{11}(q)\Big[G_{0,11}(k+q
)
G_{0,11}(k+q)
+G_{0,12}(k+q)G_{0,12}(k+q)\Big]G_{0,11}(-k)\nonumber\\
&&\times \Big[G_{0,11}(k'+q)G_{0,11}(k'+q)
+G_{0,12}(k'+q)G_{0,12}(k'+q)\Big]G_{0,11}(-k')
\nonumber\\&+&4\Gamma_{11}(q)\Gamma_{12}(q)\Big[G_{0,11}(k+q)
G_{0,11}(k+q)
+G_{0,12}(k+q)G_{0,12}(k+q)\Big]G_{0,11}(-k)\nonumber\\
&&\times 
\Big[G_{0,11}(k'+q)G_{0,12}(k'+q)
+G_{0,22}(k'+q)G_{0,12}(k'+q)\Big]G_{0,12}(-k')
\nonumber\\&+&2\Gamma_{11}(q)\Gamma_{22}(q)\Big[G_{0,11}(k+q)
G_{0,12}(k+q)
+G_{0,22}(k+q)G_{0,12}(k+q)\Big]G_{0,12}(-k)\nonumber\\
&&\times 
\Big[G_{0,11}(k'+q)G_{0,12}(k'+q)
+G_{0,22}(k'+q)G_{0,12}(k'+q)\Big]G_{0,12}(-k')
\nonumber\\
&+&2\Gamma_{12}(q)\Gamma_{12}(q)\Big[G_{0,11}(k+q)
G_{0,12}(k+q)
+G_{0,22}(k+q)G_{0,12}(k+q)\Big]G_{0,12}(-k)\nonumber\\
&&\times \Big[G_{0,11}(k'+q)G_{0,12}(k'+q)
+G_{0,22}(k'+q)G_{0,12}(k'+q)\Big]G_{0,12}(-k')
\nonumber\\&+&2\Gamma_{12}(q)\Gamma_{12}(q)\Big[G_{0,22}(k+q)
G_{0,22}(k+q)
+G_{0,12}(k+q)G_{0,12}(k+q)\Big]G_{0,22}(-k)\nonumber\\
&&\times \Big[G_{0,11}(k'+q)G_{0,11}(k'+q)
+G_{0,12}(k'+q)G_{0,12}(k'+q)\Big]G_{0,11}(-k')
\nonumber\\
&+&4\Gamma_{12}(q)\Gamma_{22}(q)\Big[G_{0,22}(k+q)
G_{0,22}(k+q)
+G_{0,12}(k+q)G_{0,12}(k+q)\Big]G_{0,22}(-k)\nonumber\\
&&\times \Big[G_{0,11}(k'+q)G_{0,12}(k'+q)
+G_{0,22}(k'+q)G_{0,12}(k'+q)\Big]G_{0,12}(-k')
\nonumber\\
&+&\Gamma_{22}(q)\Gamma_{22}(q)\Big[G_{0,22}(k+q)
G_{0,22}(k+q)
+G_{0,12}(k+q)G_{0,12}(k+q)\Big]G_{0,22}(-k)\nonumber\\
&&\times \Big[G_{0,22}(k'+q)G_{0,22}(k'+q)
+G_{0,12}(k'+q)G_{0,12}(k'+q)\Big]G_{0,22}(-k')
\Bigg\}.\label{rhonStrinati}\eea
Comparing with our results in
Eqs.~(\ref{rhosexp}) and (\ref{m11Qder1})-(\ref{m12Qder2}), it is apparent that the expression
for $\rho^B_{n,AL}$ in Eq.~(\ref{rhonStrinati}) does not contain terms
analogous to those in Eq.~(\ref{rhosexp}) arising from second-order derivatives
of $\tilde{M}_{ij}$ with respect to $Q$.  Only the terms in
Eq.~(\ref{rhosexp}) that involve products of first-order derivatives of
$\tilde{M}_{ij}$ correspond to the AL terms in Eq.~(\ref{rhonStrinati}).  
Separating the contributions from first- and second-order derivatives with 
respect to $Q$ in
Eq.~(\ref{rhosexp}), we use Eqs.~(\ref{Gamma11}) and (\ref{Gamma12}) to re-write
Eq.~(\ref{rhosexp}) as
\bea \rho^B_{n}
&=&\frac{2m}{\beta^3}\sum_q\Bigg[\Gamma_{11}(q)\Gamma_{11}(q)
\frac{\partial\tilde{M}_{11}}{\partial Q}
\frac{\partial \tilde{M}_{11}}{\partial Q} -4\Gamma_{11}(q)\Gamma_{12}(q)
\frac{\partial\tilde{M}_{11}}{\partial Q}
\frac{\partial \tilde{M}_{12}}{\partial Q}
\nonumber\\
&&+
2\Gamma_{11}(q)\Gamma_{22}(q)
\frac{\partial\tilde{M}_{12}}{\partial Q}
\frac{\partial \tilde{M}_{12}}{\partial Q}
+2\Gamma_{12}(q)\Gamma_{12}(q)
\frac{\partial\tilde{M}_{11}}{\partial Q}
\frac{\partial \tilde{M}_{22}}{\partial Q}  
\nonumber\\
&&+
2\Gamma_{12}(q)\Gamma_{12}(q)
\frac{\partial\tilde{M}_{12}}{\partial Q}
\frac{\partial \tilde{M}_{12}}{\partial Q}
-4\Gamma_{12}(q)\Gamma_{22}(q)
\frac{\partial\tilde{M}_{22}}{\partial Q}
\frac{\partial \tilde{M}_{12}}{\partial Q}+
\nonumber\\&&+\Gamma_{22}(q)\Gamma_{22}(q)
\frac{\partial\tilde{M}_{22}}{\partial Q}
\frac{\partial\tilde{M}_{22}}{\partial Q}\Bigg]
-\frac{2m^2}{\beta}\sum_q\Bigg[\frac{1}{
(M_{11}M_{22}-M_{12}M_{12})}\nonumber\\ &&\times
\left(M_{22}\frac{\partial^2\tilde{M}_{11}}{
\partial Q^2}+M_{11}\frac{\partial^2\tilde{M}_{22}}{\partial
Q^2} - 2M_{12}\frac{\partial^2\tilde{M}_{12}}{\partial
Q^2}\right)\Bigg]_{\tilde{\mu},Q\rightarrow 0}.\label{rhonED}\eea
Using Eqs.~(\ref{m11Qder1}) and (\ref{m12Qder1}), as well as
Eqs.~(\ref{G11relation}) and (\ref{G12relation}), and taking $\hat{\bQ} = 
\hat{\bz}$, one
can show that the
terms in Eq.~(\ref{rhonED}) that involve a product of
first-order $Q$-derivatives of the matrix elements $\tilde{M}_{ij}$ are
precisely equal to twice the AL expression for $\rho^B_{n,AL}$ given by
Eq.~(\ref{rhonStrinati}). To summarize, we have shown that
the expression in Eq.~(\ref{rhosexp}) reduces to Eq.~(\ref{rhonED}).  
This is equivalent to
\bea \rho^B_{n} &=&2\rho^B_{n,AL} - \frac{2m}{\beta}\sum_q\Bigg[\frac{1}{
(M_{11}M_{22}-M_{12}M_{12})}\nonumber\\ &&\times
\Bigg(M_{22}\frac{\partial^2\tilde{M}_{11}}{
\partial Q^2}+M_{11}\frac{\partial^2\tilde{M}_{22}}{\partial
Q^2}- 2M_{12}\frac{\partial^2\tilde{M}_{12}}{\partial
Q^2}\Bigg)\Bigg]_{\tilde{\mu},Q\rightarrow 0},\label{rhonED2}\eea
where $\rho^B_{n,AL}$ is given by Eq.~(\ref{rhonStrinati}).  

In
Ref.~\cite{Pieri}, it is shown that the normal fluid density is indeed twice the
AL contribution given by
Eq.~(\ref{rhonAL}), owing to an additional AL-type diagram that is
topologically non-equivalent to the diagram that gives rise to Eq.~(\ref{AL}) 
(see Fig.~2 in Ref.~\cite{Pieri}). 
This extra diagram gives rise to a contribution to the transverse current
correlation function that is equal to Eq.~(\ref{AL}) for the case of a contact interaction potential.  
Hence, the final result obtained in Ref.~\cite{Pieri} for the Bose contribution
to the transverse current
correlation function in the BCS-BEC crossover is given by
$\rho^B_{n} = 2\rho^B_{n,AL}$.

However, we see from Eq.~(\ref{rhonED2}) that in addition to the AL-type
contribution, our expression for the
normal fluid density includes terms which arise from
second-order
derivatives of the matrix elements of the inverse Gaussian fluctuation
propagator with
respect to the superfluid velocity $Q = Mv_s$.  In the BEC limit, one finds that
these
contributions vanish, which explains why Ref.~\cite{Pieri} also obtains the
Landau formula in Eq.~(\ref{rhos5}) in the
BEC limit.  However, the extra terms in
Eq.~(\ref{rhonED2}) are important in the unitarity and BCS regions.   Our
numerical results for $\rho_n$ which are discussed in Ref.~\cite{Fukushima}
include the
contributions from all terms in Eq.~(\ref{rhonED2}).  These extra terms are given explicitly by Eqs.~(\ref{m11Qder2}) and (\ref{m12Qder2}).  It would be interesting to understand which diagrams give rise to these extra contributions.


\begin{thebibliography}{99}
\bibitem{NozieresPines} P.~Nozi\`eres and D. Pines, \textit{Theory of Quantum
Liquids, Vol. II} 
(Addison-Wesley, Redwood City, Calif., 1990).
\bibitem{Khalatnikov} I.~M.~Khalatnikov, \textit{An Introduction to the Theory
of
Superfluidity} (W.~A.~Benjamin, New York, 1965).
\bibitem{Griffinbook} See, for example, A.~Griffin, \textit{Excitations in a
Bose-Condensed Liquid} (Cambridge University Press, Cambridge, 1993). 
\bibitem{NSR} P.~Nozi\`eres and S.~Schmitt-Rink, J. Low Temp. Phys. \textbf{59},
195 (1985).
\bibitem{Drummond06a} H.~Hu, X.-J.~Liu, and P.~Drummond, Phys. Rev. A 
\textbf{73}, 023617 (2006).  
\bibitem{Drummond06} H.~Hu, X.-J.~Liu, and P.~Drummond, Europhys. Lett.
\textbf{74}, 574 (2006). 
\bibitem{Astrakharchik04} G.~E.~Astrakharchik, J.~Boronat, J.~Casulleras, and
S.~Giorgini,
Phys. Rev. Lett. \textbf{93}, 200404 (2004).
\bibitem{Bulgac06} A.~Bulgac, J.~E.~Drut, and P.~Magierski, Phys. Rev. Lett.
\textbf{96}, 090404 (2006). 
\bibitem{Fisher} M.~E.~Fisher, M.~N.~Barber, and D.~Jasnow, Phys. Rev. A
\textbf{8}, 1111 (1983).
\bibitem{Engelbrecht} J.~R. Engelbrecht, M.~Randeria, and C.~A.~R. S\'a de Melo,
Phys. Rev. B  \textbf{55}, 15153 (1997).
\bibitem{Ohashi03} Y. Ohashi and A. Griffin, Phys. Rev. A \textbf{67},
063612 (2003).
\bibitem{Diener04} R.~Diener and T.-L.~Ho, cond-mat/0405174. 
\bibitem{Fukushima} N.~Fukushima, Y.~Ohashi, E.~Taylor, and A.~Griffin (to be published).
\bibitem{Astrakharchik05} G.~E.~Astrakharchik, J.~Boronat, J.~Casulleras, and
S.~Giorgini,
Phys. Rev. Lett. \textbf{95}, 230405 (2005).
\bibitem{Ceperley87} E.~L.~Pollock and D.~M.~Ceperley, Phys. Rev. B \textbf{36}, 8343 (1987).
\bibitem{Taylor05} E..~Taylor and A.~Griffin, Phys. Rev. A \textbf{72}, 053630 (2005).
\bibitem{Pieri} N.~Andrenacci, P.~Pieri, and G.~C.~Strinati, Phys. Rev. B
\textbf{68}, 144507 (2003).
\bibitem{Baym} For a review of response functions and the superfluid 
density, see G.~Baym in \textit{Mathematical Methods in Solid State and Superfluid
Theory},
edited by R.~C.~Clark and G.~H.~Derrick (Oliver and Boyd, Edinburgh, 1969).
\bibitem{Popovbook} V. N. Popov, \textit{Functional Integrals and Collective
Excitations} (Cambridge University Press, Cambridge, 1987).
\bibitem{deMelo} C.~A.~R.~ S\'a de Melo, M.~Randeria, and J.~R.~Engelbrecht,
Phys. Rev. Lett  \textbf{71}, 3202 (1993).
\bibitem{Littlewood} J.~Keeling, P.~Eastham, M.~Szymanska, and P.~Littlewood,
Phys. Rev. B  \textbf{72}, 115320 (2005).
\bibitem{YojiJPSJ} Y.~Ohashi, J. Phys. Soc. Jpn.  \textbf{74}, 2659 (2005).
\bibitem{PetrovMol} D.~S.~Petrov, C.~Salomon, and G.~V.~Shlyapnikov,
 Phys. Rev. Lett. \textbf{93}, 090404 (2004); Phys.
Rev. A \textbf{71}, 012708 (2005).
\bibitem{Ambegaokar} U.~Eckern, G.~Sch\"on, and V. Ambegaokar, Phys. Rev. B
\textbf{30}, 6419 (1984).  
\bibitem{Stone} M. Stone, Int. J. Mod. Phys. B \textbf{9}, 1359 (1995). 
\bibitem{Thouless} I. J. R. Aitchison, P. Ao, D. J. Thouless, and X.-M. Zhu,
Phys. Rev. B  \textbf{51}, 6531 (1995). 
\bibitem{QFT} To evaluate the Matsubara frequency sum leading to this
expression, see, for instance, M.~Le Bellac, \textit{Thermal Field
Theory} (Cambridge University Press, Cambridge, 1996), chap. 2.6. 
\bibitem{statphys2} E.~M.~Lifshitz and L.~P.~Pitaevskii, \textit{Statistical
Physics, Part 2} (Butterworth-Heinemann, Oxford, 2002).
\bibitem{Fetter} See, for example,  A.~L.~Fetter, Ann. Phys. (N. Y.) \textbf{60}, 464 (1970). 
\bibitem{tsuneto} E.~Abrahams and T.~Tsuneto, Phys. Rev. \textbf{152},
416 (1966).
\bibitem{stoof} H.~T.~C.~Stoof, Phys. Rev. B \textbf{47}, 7979 (1998).
\bibitem{Strinati00} P.~Pieri and G.~C.~Strinati, Phys. Rev. B \textbf{61},
15370 (2000).  
\bibitem{Strinati05} P.~Pieri, L.~Pisani, and G.~C.~Strinati, Phys. Rev. B
\textbf{72}, 012506 (2005).  
\bibitem{Brodsky06} I.~V.~Brodsky, M.~Yu.~Kagan, A.~V.~Klaptsov, R.~Combescot,
and X.~Leyronas, Phys. Rev. A \textbf{73}, 032724 (2006).
\bibitem{Heiselberg06} H.~Heiselberg, Phys. Rev. A \textbf{73}, 013607 (2006).
\end{thebibliography}
\end{document}